\documentclass[aps,prd,preprint,superscriptaddress,showpacs,nofootinbib]{revtex4-2}
\usepackage{graphicx}
\usepackage{dcolumn}
\usepackage{bm}
\usepackage{hyperref}
\usepackage[normalem]{ulem}
\usepackage{amsmath}
\usepackage{amsfonts}
\usepackage{amssymb}
\usepackage{slashed}
\usepackage{dashrule}
\usepackage[utf8]{inputenc}
\usepackage[mathscr]{euscript}
\usepackage{CJK}

\hypersetup{colorlinks=true, citecolor=blue, urlcolor=blue, linkcolor=blue}

\newcommand{\bal}{\begin{align}}
\newcommand{\eal}{\end{align}}
\newcommand{\beq}{\begin{eqnarray}}
\newcommand{\eeq}{\end{eqnarray}}
\newcommand{\nneeq}{\nonumber \end{eqnarray}}

\newcommand{\nn}{\nonumber \\}
\newcommand{\es}{& = &}
\newcommand{\rs}{\, = \,}
\newcommand{\ps}{& + &}
\newcommand{\ms}{& - &}
\newcommand{\ts}{& \times &}
\newcommand{\nt}{\nn \ts}
\newcommand{\np}{\nn \ps}
\newcommand{\nm}{\nn \ms}

\newcommand{\cD}{ {\cal D} }

\newcommand{\cE}{ {\cal E} }

\newcommand{\tdelta}{\tilde\delta}
\newcommand{\pd}{ {\partial} }
\newcommand{\ket}[1]{ {\left|{#1}\right\rangle} }
\newcommand{\bra}[1]{ {\left\langle{#1}\right|} }
\newcommand{\braket}[2]{ {\left\langle{#1}|{#2}\right\rangle} }

\newcommand{\bmat}{\left[\begin{array}}
\newcommand{\emat}{\end{array}\right]}

\begin{document}

\begin{CJK*}{UTF8}{}

	\title{All-charm tetraquark in front form dynamics}

	\CJKfamily{gbsn} 
	\author{Zhongkui Kuang (邝中奎)}
	\email{kuangzhongkui@impcas.ac.cn}
	\affiliation{Institute of Modern Physics, Chinese Academy of Sciences, Lanzhou 730000, China}
	\affiliation{CAS Key Laboratory of High Precision Nuclear Spectroscopy, Institute of Modern Physics, Chinese Academy of Sciences, Lanzhou 730000, China}
	\affiliation{School of Nuclear Science and Technology, University of Chinese Academy of Sciences, Beijing 100049, China}
	\author{Kamil Serafin}
	\email{kserafin@impcas.ac.cn}
	\affiliation{Institute of Modern Physics, Chinese Academy of Sciences, Lanzhou 730000, China}
	\affiliation{CAS Key Laboratory of High Precision Nuclear Spectroscopy, Institute of Modern Physics, Chinese Academy of Sciences, Lanzhou 730000, China}
	\author{Xingbo Zhao}
	\email{xbzhao@impcas.ac.cn}
	\affiliation{Institute of Modern Physics, Chinese Academy of Sciences, Lanzhou 730000, China}
	\affiliation{CAS Key Laboratory of High Precision Nuclear Spectroscopy, Institute of Modern Physics, Chinese Academy of Sciences, Lanzhou 730000, China}
	\affiliation{School of Nuclear Science and Technology, University of Chinese Academy of Sciences, Beijing 100049, China}
	\author{James P. Vary}
	\email{jvary@iastate.edu}
	\affiliation{Department of Physics and Astronomy, Iowa State University, Ames, Iowa 50011, USA}
	\collaboration{BLFQ Collaboration}
	\date{\today}
	
\begin{abstract}
We study all-charm tetraquarks in the front form of
Hamiltonian dynamics using the many-body basis function
approach known as basis light-front quantization.
The model Hamiltonian contains transverse and longitudinal
confining potentials and a one-gluon-exchange effective
potential. We calculate masses of two-charm-two-anticharm
states focusing on the lowest state. We also calculate
two-quark and four-quark estimates of meson-meson breakup
threshold. The results suggest that the lowest
two-charm-two-anticharm state is not a tightly
bound tetraquark. We discuss implications
of the cluster decomposition principle for theories
formulated on the light front and present our treatment
of identical particles together with color-singlet
restrictions on the space of quantum states.
\end{abstract}
\pacs{Valid PACS appear here}
\maketitle
\end{CJK*}

\section{Introduction}

Even though four-quark states, called tetraquarks, have been
studied for a long time (see Refs.~\cite{Iwasaki:1975pv,
Jaffe:1976ig,Ader:1981db,Heller:1985cb} for some early studies),
the stability of tetraquarks is still debated. One of the
basic questions is whether there exist four-quark states
whose masses are smaller than the sum of masses of two
mesons, into which the tetraquark could potentially decay
through rearrangement of quarks. Because \textit{ab initio}
calculations in QCD are challenging, researchers make use of
various strategies and approaches to estimate the masses of
tetraquarks and their results are often in conflict with each
other~\cite{Liu:2019zoy,
Lloyd:2003yc,
Barnea:2006sd,
Berezhnoy:2011xn,
Heupel:2012ua,
Bai:2016int,
Chen:2016jxd,
Karliner:2016zzc,
Wu:2016vtq,
Anwar:2017toa,
Debastiani:2017msn,
Richard:2017vry,
Esposito:2018cwh,
Richard:2018yrm,
Wang:2018poa,
Bedolla:2019zwg,
Liu:2019zuc,
Wang:2019rdo,
Albuquerque:2020hio,
Chen:2020xwe,
Deng:2020iqw,
Dong:2020nwy,
Garcilazo:2020acl,
Giron:2020wpx,
Gordillo:2020sgc,
Jin:2020jfc,
Karliner:2020dta,
Lundhammar:2020xvw,
Maciula:2020wri,
Maiani:2020pur,
Richard:2020hdw,
Sonnenschein:2020nwn,
Wang:2020dlo,
Wang:2020gmd,
Weng:2020jao,
Zhang:2020xtb}.
The goal of our paper is to initiate studies of tetraquarks
within the framework of Front Form of Hamiltonian
dynamics~\cite{Dirac:1949cp} and Basis Light-Front
Quantization (BLFQ)~\cite{Vary:2009gt}, an approach whose
ultimate goal is to achieve \textit{ab initio} calculations
in QCD. Therefore, our study is focused on the development
of the approach as much as on providing a preliminary
answer to the main tetraquark problem -- whether
or not four heavy quarks can form a bound state.

We choose to study heavy quarks (charm quarks) because
for heavy quarks one expects that the proper, QCD-based,
theoretical description can be simplified. Asymptotic
freedom, which is believed to be relevant for heavy quarks,
allows for perturbative expansion of the QCD Hamiltonian
and produces some confidence that the simple Hamiltonian
with confining and one-gluon-exchange potentials that
we use shares important features with the full QCD
Hamiltonian. Due to asymptotic freedom and quark masses
much larger than the strong interactions scale
$\Lambda_\mathrm{QCD}$, charm quarks are expected to be
relatively slow in comparison with the speed of light,
hence, additional pairs of heavy charm quarks cannot be
easily produced and should not contribute significantly
to the tetraquark dynamics. Tetraquarks of any kind
are an interesting topic of study because they are
exotic, \textit{i.e.}, they are neither mesons nor baryons,
therefore, they provide opportunities to test and extend
our understanding of hadron physics beyond the boundary
of fairly well-established meson and baryon physics.
Finally, studies of all-heavy tetraquarks received recently
additional motivation in the form of first experimental
identification of all-charm tetraquark resonance
$X(6900)$~\cite{LHCb:2020bwg}. The discovery of
a doubly charm tetraquark is also worth
noting~\cite{LHCb:2021vvq}.

BLFQ has already been used with success to study various
mesons and baryons~\cite{Li:2015zda,Li:2017mlw,Jia:2018ary,
Tang:2019gvn,Lan:2019vui,Lan:2019rba,Mondal:2019jdg,Lan:2019img,
Xu:2021wwj,Qian:2020utg} as well as in QED, see for example
Ref.~\cite{positroniumBLFQ}. However, most of those studies
involve only one Fock sector, with recently appearing
extensions~\cite{Lan:2021wok}. Questions like ``how does
confinement work?'' cannot be fully answered by studying
quark-antiquark or three-quark systems alone, even if one
uses phenomenologically successful confining potentials.
If one is to believe that gluon strings are formed in
a Hamiltonian approach to QCD (as seems to be the case for
Lattice QCD), then one is necessarily forced to explicitly
include many-gluon sectors in addition to the leading
``valence'' Fock sector. Furthermore, breaking of those
strings requires Fock sectors with additional quark-antiquark
pairs. The strength of BLFQ stems from the fact that,
in principle, it can handle many Fock sectors, each of
which can contain many particles, in a straightforward manner. 

The QCD Fock space is rich in structure and even with
the help of supercomputers the calculations are challenging,
because the dimensionality of required spaces of states
grows quickly with the addition of new Fock sectors.
The $Q Q \bar Q \bar Q$ sector is one of the natural
next targets after the $Q \bar Q$ and $Q Q Q$ sectors. 

Another important challenge resides in how to renormalize
divergent interactions of QCD. The eventual success of the
approach will probably require an adoption of effective
interactions calculated from QCD using, for example,
Renormalization Group Procedure for Effective Particles
(RGPEP)~\cite{Glazek:2012qj}. The Hamiltonian of bare,
pointlike quarks and gluons leads to the problem of
overlapping divergences~\cite{Glazek:2002jg}. RGPEP by
defining effective, finite-size particles can tame singular
interactions and reduce the number of Fock sectors
necessary to obtain satisfactory results. Effective
Hamiltonians computed using the closely related Similarity
Renormalization Group~\cite{Glazek:1993rc} (see also
Ref.~\cite{Wegner:1994}) have been successfully used in
combination with many-body methods in \textit{ab initio}
calculations in nuclear physics, see for
example~\cite{Binder:2012mk,Jurgenson:2013yya,Gebrerufael:2016xih}.
However, a relativistic quantum field theory such as QCD
is much more complicated than the non-relativistic nuclear
many-body problem of interacting nucleons.

Since we choose to deal with only charm quarks and
antiquarks we take into account antisymmetrization of
identical particles. This is also the first system treated
within BLFQ where the question about color-dependence
of the confining potential needs to be addressed because
there are two color-singlet combinations in the
$Q Q \bar Q \bar Q$ sector whereas both the $Q \bar Q$
and the $QQQ$ sectors admit only one color singlet each.
We adopt the commonly-used assumption that the confining
potential depends on color in exactly the same manner as
one-gluon-exchange interactions depend on color. We also
add a color-independent term in the longitudinal direction.
Without this added term we find some spurious, unphysical
solutions with negative mass squared.

In Sec.~\ref{sec:Hamiltonian} we present our model
many-body Hamiltonian and derive Schr\"odinger-like
equations for three cases -- describing one meson,
a tetraquark, and two mesons. The two-meson system allows
us to discuss the cluster decomposition principle on the
Light Front. Section~\ref{sec:BLFQ} is devoted to
a description of the main elements of the computational
framework of BLFQ. Our results for masses in the three
mentioned cases and a discussion about whether all-charm
tetraquarks are stable against dissociation are given
in Sec.~\ref{sec:binding}. Section~\ref{sec:conclusion}
concludes the paper. Color factors between color-singlet
states are given in the Appendix, where we describe
the procedure that takes into account Pauli exclusion
principle and allows us to work with color singlets
only.

\section{Hamiltonian}
\label{sec:Hamiltonian}

\subsection{Front Form of Hamiltonian dynamics}

Before we introduce our model Hamiltonian we mention
a few aspects of the framework we use that are important
in the context of our long-term goal of \textit{ab initio}
calculations in QCD. The Front Form of Hamiltonian
dynamics~\cite{Dirac:1949cp} has two important advantages
with respect to other Hamiltonian approaches. One of them
is the fact that particles cannot be created from free
vacuum in a way that they can be created, for example,
in the Instant Form of Hamiltonian dynamics. Front-Form
theories conserve total longitudinal momentum of particles
taking part in the interaction, where the longitudinal
momentum of a particle is defined as $p^+ = p^0 + p^3$.
In Hamiltonian approaches particles are on mass shell,
hence, $p^0 \ge |p^3|$ and $p^+$ cannot be negative.
At the same time vacuum should have $p^+ = 0$, therefore,
all particles created from vacuum should have exactly
$p^+ = 0$. Since for massive particles $p^+ \to 0$ means
energy diverging to infinity one should regularize the
theory and remove $p^+ = 0$ states, called zero modes.
However, it is also known that one cannot simply discard
those states and zero modes have to be taken into account
in some way. Even though it is an open question about
exactly what way zero modes need to be included we still
gained something: the difference between the free vacuum
and the interacting vacuum can only be contained in the
singular point $p^+ = 0$. Therefore, to a large extent
one can separate zero-modes from $p^+ > 0$ region, where
most of the usual dynamics happen [similar in form to the
Schr\"odinger equation or quark model Hamiltonians,
see Eqs.~(\ref{eq:EVEmeson}), (\ref{eq:EVEtetraquark})
and (\ref{eq:EVEtwoMesons})]. This is in contradistinction
from the Instant Form, in which particles of arbitrary
momenta can be created from the free vacuum making the
interacting vacuum a complicated state upon which one-,
two- and many-particle states are to be built.

Another advantage of the Front Form of Hamiltonian
dynamics is the fact that one can freely boost particles
and wave functions can be decomposed into products of
total and relative motion factors. This is very
fortunate because one can use exactly the same wave
functions that describe the internal structure of a hadron
regardless of how fast the hadron is moving in the
laboratory frame. Hence, the Front Form is uniquely
suited to describe high-energy processes and offers
practical advantages for building a Poincar\'e-covariant
quantum theory in a Hamiltonian approach.

In the Front Form of Hamiltonian dynamics the Hamiltonian
is $P^- = P^0 - P^3$. The momentum operators are
$P^+ = P^0 + P^3$, which is the longitudinal momentum,
and transverse momenta $P^1$ and $P^2$. We denote
two-dimensional transverse vectors with a bold font,
\textit{e.g.}, 
$\mathbf{P} = (P^1, P^2)$. The evolution of quantum
states is given by the analog of the Schr\"odinger
equation, which in the stationary version is
$P^- \ket{\Psi} = E \, \ket{\Psi}$, where $E$ is the
eigenvalue of operator $P^-$. One can also study the
closely related eigenvalue equation,
\begin{equation}
	P^\mu P_\mu \ket{\Psi} = M^2 \ket{\Psi},
	\label{eq:bound_st}
\end{equation}
where the eigenvalue $M^2$ is the invariant mass squared
of the eigenstate $\ket{\Psi}$. The eigenvalue $M^2$
depends only on relative motion of constituents and not
their absolute motion. Since we work with $P^\mu P_\mu$
instead of $P^-$ it is convenient for us to call
$H = P^\mu P_\mu$ the Hamiltonian. It is sometimes referred
to as ``light cone Hamiltonian''~\cite{Brodsky:1997de}.
Therefore,
\beq
H
\es
P^+ P^- - \mathbf{P}^2
\ .
\eeq
In the Front Form of Hamiltonian dynamics operators
$P^+$ and $\mathbf{P}$ are kinematic while $P^-$ is
dynamic. In other words, $P^-$ contains interactions,
while $P^+$ and $\mathbf{P}$ are the same regardless
of what interactions are present in the theory.

\subsection{Hamiltonian}
\label{sec:subHamiltonian}

The model Hamiltonian that we use to study four-quark
systems is
\begin{equation}
	\begin{aligned}
		H \rs H_\mathrm{kinetic} + H_\mathrm{transverse}
		+ H_\mathrm{longitudinal} + H_\mathrm{OGE} \ ,
	\end{aligned}
\label{eq:ourH}
\end{equation}
where $H_\mathrm{kinetic}$, $H_\mathrm{transverse}$,
$H_\mathrm{longitudinal}$ and $H_\mathrm{OGE}$ stand
for kinetic term, transverse confining potential term,
longitudinal confining potential term and one-gluon-exchange
(OGE) term, respectively. The kinetic energy Hamiltonian is
\begin{equation}
	\begin{aligned}
		 H_\mathrm{kinetic} \rs P^+ P_0^{-}  - \mathbf{P}^2 \ ,
	\end{aligned}
\end{equation}
where $P_0^-$ stands for the noninteracting, kinetic
part of $P^-$. The momentum operators are
\beq
P^+
\es
\int_1
p_1^+ \left(  b_1^\dag b_1 + d_1^\dag d_1 \right)
\ ,
\label{eq:Pplus}
\\
\mathbf{P}
\es
\int_1
\mathbf{p}_1 \left(  b_1^\dag b_1 + d_1^\dag d_1 \right)
\ ,
\label{eq:PplusPerp}
\\
P_0^-
\es
\int_1
p_1^- \left(  b_1^\dag b_1 + d_1^\dag d_1 \right)
\ ,
\label{eq:P0minus}
\eeq
where $p_1^- = (m^2 + \mathbf{p}_1^2)/p_1^+$, where $m$
is the quark mass and $b_1$ and $d_1$ are annihilation
operators of quark and antiquark with label $1$,
respectively.
Moreover,
\beq
\int_1
\es
\sum_{c_1, \sigma_1}
\int_0^\infty \frac{dp_1^+}{4\pi p_1^+}
\int\frac{d^2 \mathbf{p}_1}{(2\pi)^2}
\ ,
\eeq
where $c_1$ and $\sigma_1$ are the color and the
light-front helicity of particle $1$, respectively.
The normalization of operators is,
$\left\{ b_1, b_2^\dag \right\} =
\left\{ d_1, d_2^\dag \right\} = p_1^+ \tdelta_{1.2}
\delta_{\sigma_1, \sigma_2} \delta_{c_1, c_2}$,
where $\tdelta_{1.2}$ stands for the momentum
conservation Dirac delta multiplied by $16\pi^3$.

The Hamiltonians of the transverse and longitudinal
confining potentials are
\beq
H_\mathrm{transverse}
\es
\int_{121'2'} (p_1^+ + p_2^+) \, \tdelta_{12.1'2'}
\ U_{\mathrm{conf},\perp}
\ \mathrm{BD_{OGE}}
\ ,
\label{eq:Htrans}
\\
H_\mathrm{longitudinal}
\es
\int_{121'2'} (p_1^+ + p_2^+) \, \tdelta_{12.1'2'}
\ U_{\mathrm{conf},z}
\left[
a \, \mathrm{BD_{OGE}} + C_F (a-1) \, \mathrm{BD_{CI}}
\right]
\ ,
\label{eq:Hlongi}
\eeq
where $U_{\mathrm{conf},\perp}$ and $U_{\mathrm{conf},z}$
are the interaction kernels that depend on momenta and
helicities of particles $1$, $2$, $1'$, and $2'$.
The momentum conservation Dirac delta is
\beq
\tdelta_{12.1'2'}
\es
4 \pi \,  \delta\left( p_{1}^+ + p_{2}^+ - p_{1'}^+ - p_{2'}^+ \right)
\cdot (2\pi)^2 \, \delta^2 \left(
\mathbf{p}_{1} + \mathbf{p}_{2} - \mathbf{p}_{1'} - \mathbf{p}_{2'}
\right)
\ .
\eeq
The color dependence is encoded in $\mathrm{BD_{OGE}}$
and $\mathrm{BD_{CI}}$,
\beq
\mathrm{BD_{OGE}}
\es
\sum_{a = 1}^8
\left(
  \frac{1}{2} \, t^a_{11'} t^a_{22'}
  \, b_{1}^\dag b_{2}^\dag b_{2'} b_{1'}
- t^a_{11'} t^a_{2'2}
  \, b_{1}^\dag d_{2}^\dag d_{2'} b_{1'}
+ \frac{1}{2} \, t^a_{1'1} t^a_{2'2}
  \, d_{1}^\dag d_{2}^\dag d_{2'} d_{1'}
\right)
\ ,
\\
\mathrm{BD_{CI}}
\es
\delta_{c_{1},c_{1'}}
\delta_{c_{2},c_{2'}}
\left(
  \frac{1}{2}
  \, b_{1}^\dag b_{2}^\dag b_{2'} b_{1'}
+  b_{1}^\dag d_{2}^\dag d_{2'} b_{1'}
+ \frac{1}{2}
  \, d_{1}^\dag d_{2}^\dag d_{2'} d_{1'}
\right)
\ ,
\eeq
where $t^a_{ij}$ stands for $\chi_{c_i}^\dag T^a \chi_{c_j}$,
where $T^a = \frac{1}{2} \lambda^a$, with $\lambda^a$
a Gell-Mann matrix ($a = 1, 2, \dots, 8$) and $\chi_{c}
= [ \delta_{c,1}, \delta_{c,2}, \delta_{c,3} ]^T$ is
a three dimensional vector while $c = 1, 2, 3$ is the
color quantum number. In other words, $t^a_{ij}$ is half
of the matrix element of matrix $\lambda^a$ in the
$c_i$th row and $c_j$th column. The color dependence of
$\mathrm{BD_{OGE}}$ is the same as the color dependence
of the one gluon exchange, hence, the subscript ``OGE.''
On the other hand, $\mathrm{BD_{CI}}$ is diagonal in color,
color independent, hence, the subscript ``CI.'' Both
$\mathrm{BD_{OGE}}$ and $\mathrm{BD_{CI}}$ have three
terms each that describe pair-wise interactions in
quark-quark, quark-antiquark, and antiquark-antiquark
pairs. The factor $1/2$ that multiplies quark-quark
as well as antiquark-antiquark terms is present because
the two quarks, or the two antiquarks, that interact are
indistinguishable. Finally, $a$ is a constant between
$0$ and $1$, and $C_F = (N_c^2 - 1)/(2 N_c) = 4/3$ is
the value of quadratic Casimir operator in a fundamental
representation of $SU(N_c)$, $N_c = 3$. We choose
$a = 0.85$, therefore, in our Hamiltonian 85\% of
longitudinal confining strength in a meson comes
from the OGE-like term and 15\% comes from the
color-independent term. See below for more detailed
discussion.

The kernels are
\beq
U_{\mathrm{conf},\perp}
\es
  \kappa^4 
  \, \delta_{\sigma_{1}, \sigma_{1'}}
     \delta_{\sigma_{2}, \sigma_{2'}}
  \, 4\pi \, \delta(x_{12} - x_{1'2'})
  \, (x_{12} x_{21})^2
  \left[
  \frac{\pd^2}{\pd\mathbf{k}_{1'2'}^2}
  (2\pi)^2 \delta^2\left( \mathbf{k}_{1'2'} - \mathbf{k}_{12} \right)
  \right]
\ ,
\label{eq:kernelperp}
\\
U_{\mathrm{conf},z}
\es
\kappa^4 \,
\delta_{\sigma_{1}, \sigma_{1'}} \delta_{\sigma_{2}, \sigma_{2'}}
\, (2\pi)^2 \delta^2\left( \mathbf{q}_{12} - \mathbf{q}_{1'2'} \right)
\nt
\left[
-
\frac{ 1 }{ \sqrt{ D_{12} } }
\frac{ \pd }{ \pd x_{12} }
\frac{ 1 }{ \sqrt{ D_{12} } }
\frac{ 1 }{ \sqrt{ D_{1'2'} } }
\frac{ \pd }{ \pd x_{1'2'} }
\frac{ 1 }{ \sqrt{ D_{1'2'} } }
\ 4\pi \, \delta(x_{1'2'} - x_{12})
\right]
\ ,
\label{eq:kernelz}
\eeq
where $\kappa$ is the interaction strength parameter,
$x_{12} = p_1^+/(p_1^+ + p_2^+)$ is the longitudinal
momentum fraction of particle $1$ with respect to $2$
and $x_{21} = 1 - x_{12}$ is the longitudinal momentum
fraction of particle $2$ with respect to particle $1$.
Relative transverse momentum is $\mathbf{k}_{12} =
x_{21} \mathbf{p}_1 - x_{12} \mathbf{p}_2$. Moreover,
\beq
\mathbf{q}_{12}
\es
\frac{\mathbf{k}_{12}}{\sqrt{x_{12} x_{21}}}
\ ,
\label{eq:qperp}
\\
q_{12}^z
\es
m \, \frac{x_{12} - x_{21}}{\sqrt{x_{12} x_{21}}}
\ ,
\label{eq:qz}
\eeq
and
\beq
D_{12}
\rs
\frac{dq_{12}^z}{dx_{12}}(x_{12})
\es
\frac{m}{2 \, [x_{12}(1-x_{12})]^{3/2}}
\ .
\eeq
Objects with subscript $1'2'$ are defined in the same
way as objects with subscript $12$, except that $1$ is
replaced with $1'$ and $2$ is replaced with $2'$.

The confining potential is determined by the anti--de Sitter (AdS)/QCD
holography~\cite{Brodsky:2014yha} and its transverse
part reproduces the AdS/QCD harmonic oscillator in
the $Q \bar Q$ sector. In appropriate momentum
variables~\cite{Trawinski:2014msa}, in the $Q \bar Q$
sector, the longitudinal and transverse terms complement
each other and form a three-dimensional, rotationally
invariant harmonic oscillator, see
Eq.~(\ref{eq:meson}). The potentials in the $Q \bar Q$
sector are naturally extended to other sectors through
Eqs.~(\ref{eq:Htrans}) and (\ref{eq:Hlongi}), which
act in all sectors. The extension, however, is not
unique. For example, the factor $p_1^+ + p_2^+$
could be replaced with the total $P^+$. Moreover,
$\mathrm{BD_{OGE}}$ and $\mathrm{BD_{CI}}$ evaluate to
the same expression between states in the $Q \bar Q$
sector up to a factor of $C_F$. Their combination,
as in Eq.~(\ref{eq:Hlongi}), gives the result that
is independent of $a$ in the $Q \bar Q$ sector.
Our choice of the confining potential was obtained
after a study of several variants and searching for
acceptable spectral behavior of the solutions.

We found that removing color independent part or
replacing $p_1^+ + p_2^+$ with $P^+$ leads to the
appearance of unphysical solutions with negative mass
squared. While in general tachyonlike states can be
a sign of unstable equilibrium in a linear approximation
of a field theory, see Ref.~\cite{Aharonov:1969vu},
our approach is nonperturbative and we are dealing with
model Hamiltonians. Therefore, we regard the candidate
model Hamiltonians with such tachyonic solutions as unphysical.
The properties of those states are very far from
properties expected of bound tetraquark states. For
example, the dominant components of wave functions
of those non-physical states reveal very fast motion
of quarks with respect to each other making them
more like highly excited, high momentum scale states
than like states characterized by low relative momenta
appropriate to our model. $a = 0.85$ is the largest
value of $a$ that guarantees no negative $M^2$ states
appear up to $K = 50$ for $N_\mathrm{max} = 6$
(see Sec.~\ref{sec:BLFQ}). It is worth noting
that two- and more-gluon exchange potentials are in
general mixtures of OGE-like and color independent
parts. Hence, our confining potential appears reasonable,
apart from the fact that our CI potential confines
at large distances. However, the states which should
be affected the most by this confinement are the excited
states while we focus mainly on the ground state.

The Hamiltonian term of the one gluon exchange
interaction is $H_\mathrm{OGE} = P^+ V_\mathrm{OGE}$,
where
\beq
V_\mathrm{OGE}
\es
\int_{121'2'} \tdelta_{12.1'2'}
\ U_\mathrm{OGE}(1, 2; 1', 2')
\ \mathrm{BD_{OGE}}
\ .
\label{eq:interaction}
\eeq
The kernel of the OGE term is,
\beq
U_\mathrm{OGE}(1, 2; 1', 2')
\es
- g^2 \,
\frac{ \bar u_{1} \gamma_\mu u_{1'} \bar u_{2} \gamma^\mu u_{2'} }
     { (x_{12} - x_{1'2'}) \cD }
\, \sqrt{p_1^+ p_2^+ p_{1'}^+ p_{2'}^+}
\ ,
\label{eq:UOGE}
\eeq
where $\cD$ is the energy denominator,
\beq
\cal D
\es
\frac{1}{2}
  \left[
    \frac{ \mathbf{p}_1^2    + m^2 }{ x_{12} }
  - \frac{ \mathbf{p}_{1'}^2 + m^2 }{ x_{1'2'} }
  - \frac{ \mathbf{p}_2^2    + m^2 }{ x_{21} }
  + \frac{ \mathbf{p}_{2'}^2 + m^2 }{ x_{2'1'} }
  \right]
- \frac{(\mathbf{p}_1 - \mathbf{p}_{1'})^2+\mu^2}{x_{12}-x_{1'2'}}
\ ,
\eeq
with $\mu$ being a fictitious gluon mass. We use the same
spinors as those in Ref.~\cite{positroniumBLFQ} and
$ \bar u_{1} \gamma_\mu u_{1'} \bar u_{2} \gamma^\mu u_{2'} $
can be found in Table I therein. The fictitious gluon mass
$\mu$ is introduced to regulate the Coulomb singularity:
if we take $\mathbf{p}_1 = \mathbf{p}_{1'}$ and $x_{12} =
x_{1'2'}$, then $(x_{12} - x_{1'2'}) \cD$, which is in the
denominator of Eq.~(\ref{eq:UOGE}), becomes zero, unless
$\mu \neq 0$. This singularity is integrable if momenta
are continuous, however, in BLFQ we discretize longitudinal
momenta and the singularity has to be somehow regulated.
Even though diagonal matrix elements of the discretized
version of $H_\mathrm{OGE}$ diverge as $\mu \to 0$ the
eigenvalues and eigenvectors approach a finite limit.

The Hamiltonian of Eq.~(\ref{eq:ourH}) provides a unified
description of $Q \bar Q$ and $Q Q \bar Q \bar Q$ systems.
In fact, one could apply this Hamiltonian in sectors
with arbitrary number of heavy quarks and antiquarks.
We use it in three separate calculations for three purposes.
In all three cases we restrict the space of states to color
singlets, which can be achieved since $H$ conserves color.
Details are provided in the Appendix. Firstly, we solve the
$Q \bar Q$ eigenvalue problem and, by fitting the numerical
spectrum to the experimental spectrum of charmonium, we fix
the free parameters of the Hamiltonian: quark mass $m$,
confining potential strength parameter $\kappa$ and OGE
coupling constant $g$. Secondly, we solve the $Q Q \bar Q \bar Q$
eigenvalue problem to find the four-quark ground state mass.
Thirdly, we solve the $Q Q \bar Q \bar Q$ eigenvalue problem
with some interactions turned off. The interactions that are
kept allow one quark to form a meson with one antiquark and
the other quark to form a meson with the other antiquark.
There is no interaction between the two mesons and we
restrict the space of states to the states in which both
mesons are color singlets separately. This way we can
numerically estimate the two-meson threshold, which can be
different than the sum of masses of two mesons obtained
in the $Q \bar Q$ calculation due to finite basis.
Below we briefly present the three cases.

\subsection{Eigenvalue equation for mesons}
\label{sec:meson}

The Hamiltonian can have many eigenvectors of various forms.
States that describe a single meson with fixed momenta
$P_M^+$ and $\mathbf{P}_M$ are of the form,
\beq
\ket{\psi_M}
\es
\int_{12} P_M^+ \tdelta_{12.P_M} \ \psi_M(12)
\ b_1^\dag d_2^\dag \ket{0}
\ .
\eeq
The ``front-form energy'' of the meson is $P_M^- =
\frac{M^2 + \mathbf{P}_M^2}{P_M^+}$, where $M$ is the
mass of the meson. $P_M^\mu$ are eigenvalues of
operators $P^\mu$ and $M^2$ is an eigenvalue of $H$.
The eigenvalue equation $H \ket{\psi_M} = M^2 \ket{\psi_M}$
reduces to,
\beq
&&
  \frac{m^2 + \mathbf{k}_{12}^2}{x_1} \, \psi_M(12)
+ \frac{m^2 + \mathbf{k}_{12}^2}{x_2} \, \psi_M(12)
+ \sum_{c_{1'}, c_{2'}} \, \kappa^4 \, \tilde U_{12}
  \psi_{M c_{1'} c_{2'}}(12)
\nn &&
- \int_{1'2'} P_M^+ \tdelta_{1'2'.P_M}
  \, t^a_{11'} t^a_{2'2}
  \, U_\mathrm{OGE}(1,2;1',2') \, \psi_M(1'2')
\rs
M^2 \, \psi_M(12)
\ ,
\label{eq:EVEmeson}
\eeq
where
\beq
\tilde U_{12}
\es
  t^a_{11'} t^a_{2'2}
  \, x_{12} x_{21} ( \mathbf{r}_1 - \mathbf{r}_2 )^2
\nn &&
- \left[
    a \, t^a_{11'} t^a_{2'2}
  + \frac{4}{3} (1-a) \, \delta_{c_1, c_{1'}} \delta_{c_2, c_{2'}}
  \right]
  \frac{1}{\sqrt{ D_{12} }}
  \frac{\pd}{\pd x_{12}}
  \frac{1}{ D_{12} }
  \frac{\pd}{\pd x_{12}}
  \frac{1}{\sqrt{ D_{12} }}
\ .
\label{eq:U12qqbar}
\eeq
One can simplify the form of this equation
considerably by changing variables from $\mathbf{k}_{12}$
and $x_{12}$ to $\mathbf{q}_{12}$ and $q_{12}^z$
(collectively denoted $\vec{q}_{12}$) introduced in
Eqs.~(\ref{eq:qperp}) and (\ref{eq:qz}). Moreover,
we assume that the meson is a color singlet state.
Therefore,
\beq
\psi_M(12)
\es
\frac{\delta_{c_1,c_2}}{\sqrt{N_c}} \, \sqrt{D_{12}}
\, \phi_{\sigma_1 \sigma_2}(\vec{q}_{12})
\ .
\eeq
We get,
\beq
&&
  \left( 4 m^2 + \vec{q}_{12}^{\,\,2} \right)
  \phi_{\sigma_1 \sigma_2}(\vec{q}_{12})
- C_F \kappa^4
  \frac{\pd^2}{\pd\vec{q}_{12}^{\,\,2}}
  \phi_{\sigma_1 \sigma_2}(\vec{q}_{12})
\nn&&
- C_F
  \sum_{\sigma_{1'},\sigma_{2'}}
  \int\frac{d^3 q_{1'2'}}{(2 \pi)^3}
  \, \frac{ U_\mathrm{OGE}(1, 2; 1', 2') }
          { 2 \sqrt{D_{12} D_{1'2'}} }
  \, \phi_{\sigma_{1'} \sigma_{2'}}(\vec{q}_{1'2'})
\rs
M^2 \, \phi_{\sigma_1 \sigma_2}(\vec{q}_{12})
\ .
\label{eq:meson}
\eeq
This equation looks much like nonrelativistic Schr\"odinger
equation in momentum space. The Laplacian acting on the wave
function is equivalent to a rotationally symmetric harmonic
oscillator potential and the OGE potential is written in
a generic form. It is worth noting that the same confining
potential can be derived using RGPEP with a gluon mass
ansatz~\cite{Glazek:2017rwe,
Serafin:2018aih}. Our OGE
potential is different from the Coulomb plus Breit-Fermi
of Ref.~\cite{Glazek:2017rwe} and is taken instead from
Ref.~\cite{positroniumBLFQ}. The choice was dictated by
the availability of software implementation of the latter
potential. Similarly, instead of the longitudinal potential
given by Eq.~(\ref{eq:kernelz}) we could have chosen a kernel
that would give us $\pd_x x(1-x) \pd_x$ potential of
Ref.~\cite{Li:2015zda,Li:2017mlw}. In the limit of relative
momenta vanishing with respect to quark masses the two
potentials become equal, hence, both should be suitable for
phenomenology. It is sufficient for our purposes to select
one longitudinal confining potential and one OGE potential
and work with them.

\subsection{Eigenvalue equation for tetraquarks}
\label{sec:EVEtetraquark}

Tetraquark states have a form very similar to meson states,
\beq
\ket{\psi_T}
\es
\int_{1234} P_T^+ \tdelta_{1234.P_T} \ \psi_T(1234)
\ b_1^\dag b_2^\dag d_3^\dag d_4^\dag \ket{0}
\ .
\label{eq:tetraquarkState}
\eeq
This state has fixed momenta $P_T^+$ and $\mathbf{P}_T$,
while $P_T^- = \frac{M^2 + \mathbf{P}_T^2}{P_T^+}$.
The eigenvalue equation $H \ket{\psi_T} = M^2 \ket{\psi_T}$
reduces to,
\beq
&&
  \sum_{i = 1}^4
  \frac{m^2 + \mathbf{k}_i^2}{x_i} \ \psi_T(1234)
+ \sum_{i < j}
  \kappa^4 \, \tilde U_{ij} \, \psi_T(1234)
\nn&&
+ \sum_{i<j} \frac{1}{x_i + x_j}
  \int_{i'j'} (p_i^+ + p_j^+) \tdelta_{ij.i'j'}
  \, W_\mathrm{OGE}(i,j;i',j') \, \psi'
\rs
M^2 \ \psi_T(1234)
\ ,
\label{eq:EVEtetraquark}
\eeq
where $x_i = p_i^+ / P_T^+$ and $\sum_{i=1}^4 x_i = 1$,
while $\mathbf{k}_i$ is a transverse momentum of
particle $i$ in a rest frame of the bound state where
$\sum_{i=1}^4 \mathbf{k}_i = 0$. The harmonic oscillator
$\tilde U_{ij}$ for quark-antiquark interaction is given
in Eq.~(\ref{eq:U12qqbar}) with $1$, $2$, $1'$, and $2'$
replaced by $i$, $j$, $i'$, and $j'$, respectively.
For quark-quark and antiquark-antiquark it is,
\beq
\tilde U_{ij}
\es
- t^a_{ii'} t^a_{jj'}
  \, x_{ij} x_{ji} ( \mathbf{r}_i - \mathbf{r}_j )^2
\nn &&
+ \left[
    a \, t^a_{ii'} t^a_{jj'}
  - \frac{4}{3} (1-a) \, \delta_{c_i, c_{i'}} \delta_{c_j, c_{j'}}
  \right]
  \frac{1}{\sqrt{ D_{ij} }}
  \frac{\pd}{\pd x_{ij}}
  \frac{1}{ D_{ij} }
  \frac{\pd}{\pd x_{ij}}
  \frac{1}{\sqrt{ D_{ij} }}
\ .
\label{eq:U12qq}
\eeq
$W_\mathrm{OGE}(i,j;i',j') \, \psi'$ is different
depending on $i$ and $j$. For the quark-quark
interaction, \textit{i.e.}, $i=1$ and $j=2$,
\beq
W_\mathrm{OGE}(1,2;1',2') \, \psi'
\es
\frac{ t^a_{11'} t^a_{22'} \, U_\mathrm{OGE}(1,2;1',2')
     - t^a_{12'} t^a_{21'} \, U_\mathrm{OGE}(1,2;2',1') }{ 2 }
\, \psi_T(1'2'34)
\ .
\eeq
For the antiquark-antiquark interaction, $i=3$ and $j=4$,
\beq
W_\mathrm{OGE}(3,4;3',4') \, \psi'
\es
\frac{ t^a_{3'3} t^a_{4'4} \, U_\mathrm{OGE}(3,4;3',4')
     - t^a_{4'3} t^a_{3'4} \, U_\mathrm{OGE}(3,4;4',3') }{ 2 }
\, \psi_T(123'4')
\ .
\eeq
For quark-antiquark interactions, $i = 1$ or $2$ and
$j = 3$ or $4$,
\beq
W_\mathrm{OGE}(i,j;i',j') \, \psi'
\es
- t^a_{ii'} t^a_{j'j} \, U_\mathrm{OGE}(i,j;i',j') \, \psi_{i'j'}
\ ,
\eeq
where $\psi_{i'j'} = \psi_T(1'23'4)$, $\psi_T(1'234')$,
$\psi_T(12'3'4)$, and $\psi_T(12'34')$ for $ij = 13$,
$14$, $23$, and $24$, respectively. The interaction
kernels are antisymmetrized as a result of having
identical particles $b_1^\dag b_2^\dag$ and
$d_3^\dag d_4^\dag$ in Eq.~(\ref{eq:tetraquarkState}).

\subsection{Eigenvalue equation for two mesons}
\label{sec:twoMesons}

To describe two separate mesons, $A$ and $B$, we choose,
\beq
\ket{\psi_{AB}}
\es
\int_{13} P_A^+ \tdelta_{13.P_A} \,\psi_A(13)
\, b_1^\dag d_3^\dag
\int_{24} P_B^+ \tdelta_{24.P_B} \,\psi_B(24)
\, b_2^\dag d_4^\dag \ \ket{0}
\ .
\eeq
Meson $A$ has momentum components $P_A^+$ and $\mathbf{P}_A$,
while meson $B$ has momentum components $P_B^+$ and
$\mathbf{P}_B$. By placing the two mesons far enough
from each other we can make the total interaction between
them to be arbitrarily small. We simulate this situation
by turning off all interactions except those between
particles $1$ and $3$, which form meson $A$, and between
particle $2$ and $4$, which form meson $B$. Moreover, two
identical quarks contained in two separated mesons are in
practice distinguishable. Therefore, in this section we
treat all particles as distinguishable. Since there are
no interactions between the two mesons, we expect that
in the general eigenvalue equation,
\beq
H \, \ket{\psi_{AB}}
\es
M^2 \, \ket{\psi_{AB}}
\ ,
\label{eq:EVE}
\eeq
the eigenvalue $M^2$ can be written as the invariant
mass of two mesons with mass $M_A$ and $M_B$,
\beq
M^2
\es
(P_{A\mu} + P_{B\mu}) (P_A^\mu + P_B^\mu)
\rs
  \frac{M_A^2 + \mathbf{k}_{AB}^2}{x_A}
+ \frac{M_B^2 + \mathbf{k}_{AB}^2}{x_B}
\ .
\label{eq:composedMasses}
\eeq
The relative transverse momentum between mesons is
$\mathbf{k}_{AB} = x_B \mathbf{P}_A - x_A \mathbf{P}_B$,
where $x_A = P_A^+ / (P_A^+ + P_B^+) = x_1 + x_3$,
$x_B = x_2 + x_4$. Equation~(\ref{eq:EVE}) reduces to,
\beq
  \frac{1}{x_A} \cE_A \psi_B(24)
+ \frac{1}{x_B} \cE_B \psi_A(13)
\es
\left(
  \frac{M_A^2}{x_A}
+ \frac{M_B^2}{x_B}
\right) \psi_A(13) \, \psi_B(24)
\ ,
\label{eq:EVEtwoMesons}
\eeq
where
\beq
\cE_A
\es
  \frac{m^2 + \mathbf{k}_{13}^2}{x_{13} x_{31}} \, \psi_A(13)
- x_A \sum_{c_{1'}, c_{3'}} \kappa^4 \, \tilde U_{13}
  \psi_{A c_{1'}c_{3'}}(13)
\nm
  \int_{1'3'} P_A^+ \tdelta_{1'3'.P_A}
  \, t_{11'}^a t_{3'3}^a
  \, U_\mathrm{OGE}(1, 3; 1', 3') \, \psi_A(1'3')
\ ,
\\
\cE_B
\es
  \frac{m^2 + \mathbf{k}_{24}^2}{x_{24} x_{42}} \, \psi_B(24)
- x_B \sum_{c_{2'}, c_{4'}} \kappa^4 \, \tilde U_{24}
  \psi_{B c_{2'} c_{4'}}(24)
\nm
  \int_{2'4'} P_B^+ \tdelta_{2'4'.P_B}
  \, t_{22'}^a t_{4'4}^a
  \, U_\mathrm{OGE}(2, 4; 2', 4') \, \psi_B(2'4')
\ .
\eeq
Note, that the relative transverse kinetic energy between
the mesons in the eigenvalue, $\mathbf{k}_{AB}^2 / x_A
+ \mathbf{k}_{AB}^2 / x_B$, canceled with the transverse
kinetic energy between mesons in $H_\mathrm{kinetic}$,
which is fixed by the choice of state $\ket{\psi_{AB}}$.
We separate Eq.~(\ref{eq:EVEtwoMesons}) into two,
$\cE_A = M_A^2 \psi_A(13)$ and $\cE_B = M_B^2 \psi_B(24)$.
Using the same kind of substitution as in Sec.~\ref{sec:meson},
\beq
\psi_A(13)
\es
\frac{\delta_{c_1,c_3}}{\sqrt{N_c}} \, \sqrt{D_{13}}
\, \phi_{A \sigma_1 \sigma_3}(\vec{q}_{13})
\ ,
\\
\psi_B(24)
\es
\frac{\delta_{c_2,c_4}}{\sqrt{N_c}} \, \sqrt{D_{24}}
\, \phi_{B \sigma_2 \sigma_4}(\vec{q}_{24})
\ ,
\eeq
we get two eigenvalue equations,
\beq
&&
  \left( 4 m^2 + \vec{q}_{13}^{\,\,2} \right)
  \phi_{A \sigma_1 \sigma_3}(\vec{q}_{13})
- x_A C_F \kappa^4
  \frac{\pd^2}{\pd\vec{q}_{13}^{\,\,2}}
  \phi_{A \sigma_1 \sigma_3}(\vec{q}_{13})
\nn&&
- C_F
  \sum_{\sigma_{1'},\sigma_{3'}}
  \int\frac{d^3 q_{1'3'}}{(2 \pi)^3}
  \, \frac{ U_\mathrm{OGE}(1, 3; 1', 3') }
          { 2 \sqrt{D_{13} D_{1'3'}} }
  \, \phi_{A \sigma_{1'} \sigma_{3'}}(\vec{q}_{1'3'})
\rs
M_A^2 \, \phi_{A \sigma_1 \sigma_3}(\vec{q}_{13})
\ ,
\label{eq:mesonA}
\\
&&
  \left( 4 m^2 + \vec{q}_{24}^{\,\,2} \right)
  \, \phi_{B \sigma_2 \sigma_4}(\vec{q}_{24})
- x_B C_F \kappa^4
  \frac{\pd^2}{\pd\vec{q}_{24}^{\,\,2}}
  \phi_{B \sigma_2 \sigma_4}(\vec{q}_{24})
\nn&&
- C_F
  \sum_{\sigma_{2'},\sigma_{4'}}
  \int\frac{d^3 q_{2'4'}}{(2 \pi)^3}
  \, \frac{ U_\mathrm{OGE}(2, 4; 2', 4') }
          { 2 \sqrt{D_{24} D_{2'4'}} }
  \, \phi_{B \sigma_{2'} \sigma_{4'}}(\vec{q}_{2'4'})
\rs
M_B^2 \, \phi_{B \sigma_2 \sigma_4}(\vec{q}_{24})
\ .
\label{eq:mesonB}
\eeq

\subsection{Cluster decomposition principle}
\label{sec:clusterDP}

The two-meson solutions in the $Q Q \bar Q \bar Q$
sector provide a good example of how the cluster
decomposition principle works in the Front Form
of Hamiltonian dynamics. There are several elements
needed for the cluster decomposition principle to be
satisfied. First of all, the mass $M_A$ of meson $A$
should not depend on the state of particles in
meson $B$. Similarly, the mass $M_B$ of meson $B$
should not depend on the state of particles in
meson $A$. Secondly, $M_A$ and $M_B$ calculated
in the $Q Q \bar Q \bar Q$ sector should be equal
to the corresponding masses in the $Q \bar Q$ sector.
For example, if meson $A$ is in the $0^{-+}$ ground
state and meson $B$ is in the $1^{--}$ ground state,
then $M_A$ should be exactly equal to the mass of
$\eta_c$ calculated in the $Q \bar Q$ sector and
$M_B$ should be exactly equal to the mass of $J/\psi$
calculated in the $Q \bar Q$ sector. That is expected
from the analytic solutions, numerical solutions
may differ slightly.

In the two-meson example in Sec.~\ref{sec:twoMesons}
those conditions are not satisfied. Comparing
Eq.~(\ref{eq:mesonA}) with Eq.~(\ref{eq:meson}) one
can see that in Eq.~(\ref{eq:mesonA}) there is an extra
factor $x_A$ multiplying the confining potential. Since
$x_A$ is fixed the mass $M_A$ is independent of whether
meson $B$ is in the $\eta_c$ or $J/\psi$ or any other
state. Nevertheless, $M_A$ does depend on $P_B^+$
because $x_A$ depends on $P_B^+$. Moreover, $M_A$ cannot
be the same as the mass of the corresponding charmonium
in the $Q \bar Q$ sector, because the strength of the
confining potential in the $Q \bar Q$ sector is
$C_F \kappa^4$, while it is $x_A C_F \kappa^4$ for
meson $A$ in the $Q Q \bar Q \bar Q$ sector.

We could formally restore the decomposition principle
by replacing $p_1^+ + p_2^+$ in Eqs.~(\ref{eq:Htrans})
and (\ref{eq:Hlongi}) with $P^+$, but this would lead
to the appearance of spurious states as described in
Sec.~\ref{sec:subHamiltonian}. We prioritize the
acceptable spectrum over exact conservation of the
decomposition principle, since the former is more
important in practice, while the latter can be
approximately restored. Since charm quarks are heavy, the
two-meson system and tetraquark can be considered as
nonrelativistic. Therefore, $x_A \approx x_B \approx 1/2$
and one can partly restore the cluster decomposition
principle by rescaling $\kappa$ in the $Q Q \bar Q \bar Q$
sector. In other words, in the $Q Q \bar Q \bar Q$
sector we use $\kappa_T = 2^{1/4} \kappa$ instead
of $\kappa$. This guarantees that $x_A C_F \kappa_T^4
\approx C_F \kappa^4$ for $x_A \approx 1/2$.

As opposed to the confining potential, the OGE potential
fully obeys the cluster decomposition principle.
This is to be expected because it can be derived
from QCD in perturbation theory. In fact, all two-body
potentials in QCD have the same generic form of
Eq.~(\ref{eq:interaction}) (apart from the color
factors, which may differ). It is important that
$U_\mathrm{OGE}(1, 2; 1', 2')$ depends only on relative
momenta between particles $1$ and $2$ and between
$1'$ and $2'$, and not on, \textit{e.g.}, momentum
fractions $x_1$ or $x_2$, which depend on the total
$P^+$ of the system. Therefore, Eq.~(\ref{eq:interaction})
illustrates the general form of two-body operators
that admit the cluster decomposition principle in the
sense described here. A more general treatment of
relativistic theories obeying cluster separability
can be found in Ref.~\cite{Coester:1982vt}.

In the eigenvalue equations the cluster decomposition
principle manifests itself by the presence of $1/x_A$
and $1/x_B$ factors in Eq.~(\ref{eq:EVEtwoMesons}) and
the $1/(x_i + x_j)$ factor in Eq.~(\ref{eq:EVEtetraquark})
that multiply interaction terms that depend only on
relative momenta within the interacting pair with no
trace of the total $P^+$. Note that momentum conservation
in $\int_{i'j'} (p_i^+ + p_j^+) \tdelta_{ij.i'j'}$
fixes $p_{i'} + p_{j'}$ and one is left with an integral
over relative momenta $x_{i'j'}$ and $\mathbf{k}_{i'j'}$.

\section{Basis Light-Front Quantization And Truncation Scheme}
\label{sec:BLFQ}

Basis Light Front Quantization is a basis function approach
to Hamiltonian light-front field theories~\cite{Vary:2009gt}.
Longitudinal and transverse directions are treated differently.
In the longitudinal direction a box of length $2L$,
\textit{i.e.}, coordinate $x^- \in [-L, L]$, is introduced.
This leads to discretization of the longitudinal momenta.
We apply antiperiodic boundary condition for the quark field,
which means that quark longitudinal momenta can only take
values,
\beq
p^+ \es \frac{2\pi}{L} k
\ ,
\eeq
where $k$ is called the longitudinal quantum number and
it is a positive half-integer. In sectors with many
particles the total longitudinal momentum is by definition
$P^+ = \frac{2\pi}{L} K$, where $K = \sum_i k_i$ is the sum
of longitudinal quantum numbers of all particles. In
Sec.~\ref{sec:Hamiltonian} $P^+$ denoted the momentum
operator, from now on $P^+$ means the eigenvalue of the
operator $P^+$ and we keep it fixed (we use only eigenstates
of the operator $P^+$ with eigenvalue $P^+$). For a given
particle $i$, the longitudinal momentum fraction $x_i$ is,
\beq
x_i \es \frac{p_i^+}{P^+} \rs \frac{k_i}{K}
\ .
\eeq
The longitudinal continuum limit is $L, K \to \infty$
while keeping $P^+$ fixed. None of the quantities that
we calculate depend on the exact values of $P^+$ and
$L$ due to Front-Form boost invariance.

For transverse momenta we introduce the harmonic
oscillator basis~\cite{Vary:2009gt}. We define new
creation and annihilation operators,
\beq
B_i
\es
\frac{1}{\sqrt{P^+}}
\int\frac{d^2q}{(2\pi)^2} \, \Psi_{n_i}^{m_i}(\mathbf{q})^*
\left. b_i \right|_{\mathbf{p}_i = \sqrt{x_i} \mathbf{q}}
\ ,
\label{eq:operatorB}
\\
D_i
\es
\frac{1}{\sqrt{P^+}}
\int\frac{d^2q}{(2\pi)^2} \, \Psi_{n_i}^{m_i}(\mathbf{q})^*
\left. d_i \right|_{\mathbf{p}_i = \sqrt{x_i} \mathbf{q}}
\ .
\label{eq:operatorD}
\eeq
Note that the operators $B_i$ and $D_i$
depend on discrete quantum numbers $n_i$, $m_i$, $k_i$,
$\sigma_i$ and $c_i$, while plane-wave operators $b_i$
and $d_i$ depend on continuum transverse momentum
$\mathbf{p}_i = \sqrt{x_i} \mathbf{q}$, discretized
longitudinal momentum $p_i^+ = 2\pi k_i / L$ (or
equivalently on $k_i$), and on spin and color $\sigma_i$
and $c_i$. Operators $B_i$ and $D_i$ are normalized
to unity, that is,
\beq
\left\{ B_i, B_j^\dag \right\}
\rs
\left\{ D_i, D_j^\dag \right\}
\es
   \delta_{n_i, n_j}
\, \delta_{m_i, m_j}
\, \delta_{k_i, k_j}
\, \delta_{\sigma_i, \sigma_j}
\, \delta_{c_i, c_j}
\ .
\eeq
The basis wave functions are,
\beq
\Psi_n^m(\mathbf{q})
\es
\frac{1}{b}
\sqrt{\frac{4\pi n!}{(n+|m|)!}}
\, L_n^{|m|}\left( \frac{q^2}{b^2} \right)
e^{-\frac{q^2}{2b^2}}
\left|\frac{q}{b}\right|^{|m|} e^{im\varphi}
\ ,
\label{eq:HOwaveFunctions}
\eeq
where $L_n^{|m|}$ are the associated Laguerre polynomials,
$q = \sqrt{(q^1)^2 + (q^2)^2}$, $\varphi = \arg\mathbf{q}$,
and $b$ is a selectable positive parameter of dimension
of $\mathbf{P}$. The principal quantum number $n$ is a
non-negative integer, while $m$ can be an arbitrary integer.
The choice of the harmonic-oscillator wave functions is
compatible with our choice of the transverse confining
potential and is important for the factorization of the
center-of-mass motion, which we describe in detail later
in this section.

In practice one has to truncate the many-particle basis
in the transverse direction by limiting the allowed radial
numbers $n_i$ and angular numbers $m_i$ by a cutoff in the
number of oscillator quanta in each basis state,
\beq
\sum_i \left( 2 n_i + |m_i| + 1 \right)
& \leq &
N_\mathrm{max}
\ .
\label{eq:truncationNmax}
\eeq
Removing this truncation is equivalent with taking the
limit $N_\mathrm{max} \to \infty$. In addition, we
require our multi-particle basis state to have
total angular momentum projection,
\beq
M_J \es \sum_i\left(m_i + \sigma_i\right)
\ ,
\eeq
where $\sigma_i = \pm\frac{1}{2}$ is the fermion light-front
helicity. Throughout this article we limit our attention to
$M_J = 0$ states.\footnote{A tetraquark state with
$M_J = 0$ can, in principle, be built from one meson having,
for example, $M_J = +1$ and the other having $M_J = -1$.
However, this is not expected to play a role in calculations
focused on the tetraquark ground state since such states
would be expected to result in a higher tetraquark
dissociation threshold than the one where both mesons have
$M_J=0$.} It is also worth mentioning that the truncation
of the basis breaks the cluster decomposition principle.
For example, if we consider our two-meson example from
Sec.~\ref{sec:twoMesons} and if the quantum numbers of
particles forming meson $A$ already almost saturate
Eq.~(\ref{eq:truncationNmax}), then the particles of meson
$B$ will be restriced to a much smaller space of states
than the particles of meson $A$. The opposite situation
is also possible and included in the truncated basis.
Therefore, one meson can influence the other through the
truncation, even if there are no interactions between them.
Moreover, each of the mesons in the $Q Q \bar Q \bar Q$
sector is subject to a different truncation than the one
meson in the $Q \bar Q$ sector and meson masses in the
$Q \bar Q$ and in the $Q Q \bar Q \bar Q$ sectors can
differ slightly, but the difference should vanish as
the basis size is increased.

It is straightforward to rewrite the Hamiltonian presented in
Sec.~\ref{sec:Hamiltonian} using new operators $B$ and $D$.
One has to additionally discretize the longitudinal momenta
by following a simple prescription,
$4\pi\delta(p_1^+ - p_2^+) \to 2 L \delta_{k_1, k_2}$,
$\int_0^\infty\frac{dp^+}{4\pi} \to \frac{1}{2L} \sum_k$,
$b \to \sqrt{2L} b$, $d \to \sqrt{2L} d$. Then it is
only a matter of computing matrix elements of $H$ and
diagonalizing the obtained matrix to obtain eigenstates of
$H$ and their masses. Computation of matrix elements between
states containing two quarks and two antiquarks is not much
more complicated than the analogous computation between
states containing only one quark and one antiquark because
no particle nor any pair of particles is distinguished. One,
obviously, has to calculate terms for all six pairs of
particles instead of just one, and interactions between
identical particles must be property antisymmetrized. Using
a basis in relative momenta of Jacobi type, for example,
would require us to use different formulas for different
pairs of interacting particles. It should be evident that
the addition of more particles in our calculation (including
gluons) would be straightforward. Admittedly, this comes
at a cost of larger matrices (effectively one more particle
per Fock sector compared to Jacobi coordinates), but the
larger matrices are also more sparse which aids applications
on modern computers, while the simplicity makes the software
development more reliable. Probably the most important
complication is introduced by restricting our space of
states to only color-singlet states. This important, but
rather technical topic is described in more detail in
the Appendix. Similar basis spaces restricted to include
only color singlets have been implemented for a BLFQ
treatment of glueballs with Fock spaces having up to six
gluons~\cite{Vary:2018pmv}.

Since BLFQ implements states using single-particle transverse
motion instead of relative motion, the resulting eigenvectors
will possess center-of-mass motion excitations which are of
no interest to us because they do not influence the invariant
mass nor the internal structure of hadrons on the light front.
The harmonic oscillator basis allows us to easily deal with
this problem. By adopting Eq.~(\ref{eq:truncationNmax}) the
eigenvectors of the truncated Hamiltonian have a known and simple
center-of-mass motion. This can be demonstrated by showing
that, even in the truncated basis, $H$ commutes with the similarly
truncated center-of-mass harmonic-oscillator Hamiltonian
\beq
H_\mathrm{CM}
\es
\lambda_\mathrm{CM}
\left( \mathbf{P}^2 + b^4 \mathbf{R}^2 - 2 b^2 \right)
\ ,
\eeq
where $\mathbf{P}$ is the transverse momentum operator
and $\mathbf{R}$ is the transverse center-of-mass position
operator,
\beq
\mathbf{R}
\es
\sum_{k_1,k_2,\sigma_1,\sigma_2,c_1,c_2}
\frac{
\delta_{k_1,k_2}
\, \delta_{\sigma_1,\sigma_2}
\, \delta_{c_1,c_2}
}{P^+}
\int\frac{d^2 \mathbf{p}_1}{(2\pi)^2}
\int\frac{d^2 \mathbf{p}_2}{(2\pi)^2}
\nt
\left[ -i \frac{\pd}{\pd\mathbf{p}_1} (2\pi)^2
\delta^2(\mathbf{p}_1 - \mathbf{p}_2)
\right]
\left( b_1^\dag b_2 + d_1^\dag d_2 \right)
\ .
\eeq
Eigenvalues of $H_\mathrm{CM}$ are $n \cdot 2 b^2
\lambda_\mathrm{CM}$, where $n$ is a non-negative integer.
$n = 0$ corresponds to the ground state of center-of-mass
motion and $n \ge 1$ correspond to excited states of
center-of-mass motion. In a typical scenario among
eigenstates of $H$ with the lowest eigenvalues there
will be states with the same relative motion but different
center-of-mass motion. To keep only the eigenstates with
the ground-state center-of-mass motion we diagonalize
$H + H_\mathrm{CM}$ instead of $H$. Since $H$ and
$H_\mathrm{CM}$ commute they have the same eigenvectors,
while the eigenvalues of the sum will be the sum of the
eigenvalues of $H$ and $H_\mathrm{CM}$. Therefore, states
with excited-state center-of-mass motion will be shifted
up by a multiple of $2 b^2 \lambda_\mathrm{CM}$. Choosing
$\lambda_\mathrm{CM}$ sufficiently large and positive,
all states with excited center-of-mass motion will have
eigenvalues larger than the eigenvalues of the limited
number of states that we obtain numerically. We use
$\lambda_\mathrm{CM} = 50$ in our calculations.

\section{Analysis of binding energy}
\label{sec:binding}

To address the question whether there exist 
$c c \bar c \bar c$ states that cannot break up into
two charmonia we need to know the mass of the lowest
tetraquark state and the value of the two-charmonium
threshold taking into account the implications of the
truncated basis space for the subsystems. We obtain
estimates of both by numerically diagonalizing truncated
matrices of our model Hamiltonian obtained using BLFQ.
We therefore solve three problems which correspond to
three eigenvalue equations presented in Secs.~\ref{sec:meson},
\ref{sec:EVEtetraquark} and \ref{sec:twoMesons}.
The discrete spectra of truncated Hamiltonians should
look more and more like the spectrum of the untruncated,
infinite Hamiltonian as $N_\mathrm{max} \to \infty$
and $K \to \infty$.

The Hamiltonian matrix in the sector with one meson
is used to fix free parameters of the model, $m$, $\kappa$
and $\alpha = g^2/(4\pi)$. The gluon mass $\mu$ = 10~MeV
and the basis parameter $b$ in the meson calculation is
fixed to be equal to $\kappa$. We fit the lowest eight
states in the spectrum of charmonium. The root mean square
difference between fitted and experimental masses is 31~MeV.
The parameters are given in Table~\ref{tab:parameters}, while
Table~\ref{tab:mesonMasses} lists the fitted meson masses
and the corresponding experimental values. The fitting was
carried out for $N_\mathrm{max} = 6$ and $K = 9$. In all
calculations we calculate $M_J = 0$ states. For the purpose
of estimating the two-meson threshold, see
Eq.~(\ref{eq:threshold1}), we calculated also meson masses
for all $2 \le N_\mathrm{max} \le 10$ and $1 \le K \le 17$,
and $c \bar c$ masses with interactions between $c$ and
$\bar c$ turned off for the same range of $N_\mathrm{max}$
and $K$. The lowest possible $N_\mathrm{max}$ in a system
with two particles is 2 while the lowest $K$ is 1. The
upper bounds on $N_\mathrm{max}$ and $K$ for a meson are
determined by the largest $N_\mathrm{max}$ and $K$
that we used in tetraquark computations.
\begin{table}[h]
\begin{ruledtabular}
\caption{Parameters obtained from fit to experimental
meson spectrum.
\label{tab:parameters}}
\begin{tabular}{ccc}
$m$ & $\kappa$ & $\alpha$ \\
\hline
1.25~GeV & 1.21~GeV & 0.367
\end{tabular}
\end{ruledtabular}
\end{table}
\begin{table}[h]
\begin{ruledtabular}
\caption{Fitted masses in MeV. $N_\mathrm{max} = 6$, $K = 9$.
\label{tab:mesonMasses}}
\begin{tabular}{ccccccccc}
&
$\eta_c(1S)$ & $J/\psi$ & $\chi_{c0}$ & $\chi_{c1}$ &
$\chi_{c2}$ & $h_c$ &  $\eta_c(2S)$ & $\psi(2S)$
\\
\hline
Fit &
3031 & 3067 & 3415 & 3517 & 3564 & 3474 & 3676 & 3666
\\
Exp. &
2984 & 3097 & 3415 & 3511 & 3556 & 3525 & 3637 & 3686
\end{tabular}
\end{ruledtabular}
\end{table}

Tetraquark masses are calculated for $N_\mathrm{max}$ =
6, 8, 10, 12 and for $K$ = 6, 10, 14, 18. We calculated
three sets of tetraquark masses: masses with all interactions
turned on, \textit{cf.} Eq.~(\ref{eq:EVEtetraquark}), masses
with all interactions turned off, $\kappa = 0$, $\alpha = 0$,
and masses with only interactions in the pair $13$ and in
the pair $24$ turned on, \textit{cf.}
Eq.~(\ref{eq:EVEtwoMesons}).
For the purpose of the tetraquark calculations we readjust
the parameter $b$ to remove possible source of mismatch
between meson and tetraquark calculations. The characteristic
value of $q$ in Eq.~(\ref{eq:HOwaveFunctions}) is $b$, which
means that the characteristic value of $|\mathbf{p}|$ is
$\sqrt{x} b$. In the tetraquark case we have similarly,
$|\mathbf{p}| \sim \sqrt{x} b$. However, in the meson case,
the expected value of $x$ is $1/2$, while in the tetraquark
case the expected value of $x$ is $1/4$. Therefore, it is
reasonable to take $b_T = \sqrt{2} b_M$, where $b_T$ is the
value of $b$ for tetraquark calculations and $b_M$ is
the value of $b$ for meson calculations. This way the
characteristic scale of $|\mathbf{p}|$ in the basis is the
same in the two cases. This readjustment is not strictly
necessary, because for sufficiently large $N_\mathrm{max}$
and $K$ the results should be rather insensitive to the
choice of $b$ for fixed $\kappa$ over a wide range of
values of $b$, but it should increase the utility of the
results for small $N_\mathrm{max}$ and $K$. Moreover, as
already mentioned in Sec.~\ref{sec:clusterDP}, instead of
$\kappa$ we use $\kappa_T = 2^{1/4} \kappa$ for the
confining strength parameter.

One of the sources of systematic errors of the 
framework we adopt originates from the fact that a pair
of particles in $c c \bar c \bar c$ system has a minimal
nonzero kinetic energy with respect to the other two particles.
The minimal kinetic energy should approach zero from above
as $N_\mathrm{max}$ approaches infinity, but may be of
importance for finite $N_\mathrm{max}$. This artifact of
a finite basis is called ``kinetic energy penalty''
in Ref.~\cite{Lloyd:2003yc}. Here we estimate it in the
following way,
\beq
\Delta M^2
\es
  M_{c c \bar c \bar c}^{2\,\mathrm{free}}(N_\mathrm{max},K)
-
 \min_{N_1,K_1}\left[
    \frac{ M^{2\,\mathrm{free}}_{c \bar c}(N_1,K_1) }{ K_1/K }
  + \frac{ M^{2\,\mathrm{free}}_{c \bar c}(N_2,K_2) }{ K_2/K }
  \right]
\ ,
\label{eq:deltaMfreeTetra}
\eeq
where $N_2$ and $K_2$ are fixed by conditions
$N_1 + N_2 = N_\mathrm{max}$ and $K_1 + K_2 = K$.
$M_{c c \bar c \bar c}^{2\,\mathrm{free}}$ and
$M^{2\,\mathrm{free}}_{c \bar c}$ are tetraquark and
meson ground-state masses squared, respectively,
computed with all interactions turned off. We use
masses squared instead of masses because they, and not
the masses, are the eigenvalues of our Hamiltonian.
Moreover, the two two-quark masses,
$M^{2\,\mathrm{free}}_{c \bar c}(N_1,K_1)$ and
$M^{2\,\mathrm{free}}_{c \bar c}(N_2,K_2)$, need
to be combined
according to Eq.~(\ref{eq:composedMasses}) to get the
invariant mass of the full state. To get the minimal
invariant mass we put $\mathbf{k}_{AB} = 0$ and minimize
over all possible values of $N_1$, $N_2$ and $K_1$, $K_2$
into which $N_\mathrm{max}$ and $K$ can be partitioned.
Hence, $\Delta M^2$, being the difference between
the actual tetraquark mass and the minimal possible
mass of two separate two-quark subsystems, is a measure
of minimal $\mathbf{k}_{AB}$ between the two subsystems.
Table~\ref{tab:deltaMfreeTetra} lists the values we
obtain. Note, that $\Delta M^2$ does not depend on $K$
for the choice of $K$s we made. It should stay the same
for all $K = 2$ (mod 4). We correct the actual eigenvalues
of the truncated Hamiltonians by subtracting the kinetic
energy penalty,
\beq
M^{2\,\mathrm{corrected}}_{c c \bar c \bar c}
\es
  M^{2\,\mathrm{full}}_{c c \bar c \bar c}(N_\mathrm{max},K)
- \Delta M^2(N_\mathrm{max},K)
\ .
\eeq
To give an estimate for a typical downward shift of
tetraquark masses introduced by this correction, for
$N_\mathrm{max}$ = 12, if
$M^\mathrm{full}_{c c \bar c \bar c}$ = 6~GeV, then
$M^\mathrm{full}_{c c \bar c \bar c}
- M^\mathrm{corrected}_{c c \bar c \bar c} \approx$ 49~MeV.
\begin{table}[h]
	\begin{ruledtabular}
		\caption{Kinetic energy penalty in GeV$^2$.
		\label{tab:deltaMfreeTetra}}
		\begin{tabular}{ccccc}
			$N_\mathrm{max}$ & 6 & 8 & 10 & 12 \\
			\hline
			$\Delta M^2$ &
			$1.213$ & $1.013$ & $0.659$ & $0.584$
		\end{tabular}
	\end{ruledtabular}
\end{table}

We introduce three estimates of the threshold with
which we compare our numerical tetraquark masses.
One estimate uses the same idea behind the second
term in Eq.~(\ref{eq:deltaMfreeTetra}) but with
full meson masses that include interactions,
\beq
T_1'
\es
\sqrt{
\min_{N_1,K_1}\left[
  \frac{ M^{2\,\mathrm{full}}_{c \bar c}(N_1,K_1) }{ K_1/K }
+ \frac{ M^{2\,\mathrm{full}}_{c \bar c}(N_2,K_2) }{ K_2/K }
\right]
}
\ ,
\label{eq:threshold1}
\eeq
where the minimum, just like in Eq.~(\ref{eq:deltaMfreeTetra}),
is taken over all possible values of $N_1$ and $K_1$.
Threshold $T_1'$ gives an unexpectedly poor estimate.
It is substantially smaller than twice our fitted
numerical mass of $\eta_c$. The reason seems to be
overestimation of the OGE potential for small values
of $K$ because the minima of $T_1'$ tend to be reached
at the minimal $K_1 = 1$, while turning off OGE potential
makes the minima to appear for $K_1 = K_2 = K/2$. In fact,
one naively expects the minimum in the definition of $T_1'$
to be reached for $K_1 = K/2$, because it implies
$x_A = x_B = 1/2$, which means zero relative longitudinal
momentum between the two mesons (as long as they have equal
masses). Moreover, the actual $M^{2\,\mathrm{full}}_{c \bar c}$
turns out to be negative for some $K$ = 1 cases, which
is unacceptable. Therefore, we define another estimate
of the threshold, for which both $N_\mathrm{max}$ and
$K$ are equally partitioned among $N_1$, $N_2$ and
$K_1$, $K_2$, \textit{i.e.},
\beq
T_1(N_\mathrm{max},K)
\es
2 \sqrt{ M^{2\,\mathrm{full}}_{c \bar c}
\left( \frac{N_\mathrm{max}}{2},\frac{K}{2} \right) }
\ .
\label{eq:threshold1p}
\eeq
This estimate seems to be more reasonable and it is in
rough agreement with a third estimate of the threshold
provided below. 

\begin{figure}[h]
 \centering
 \includegraphics[width=\linewidth]{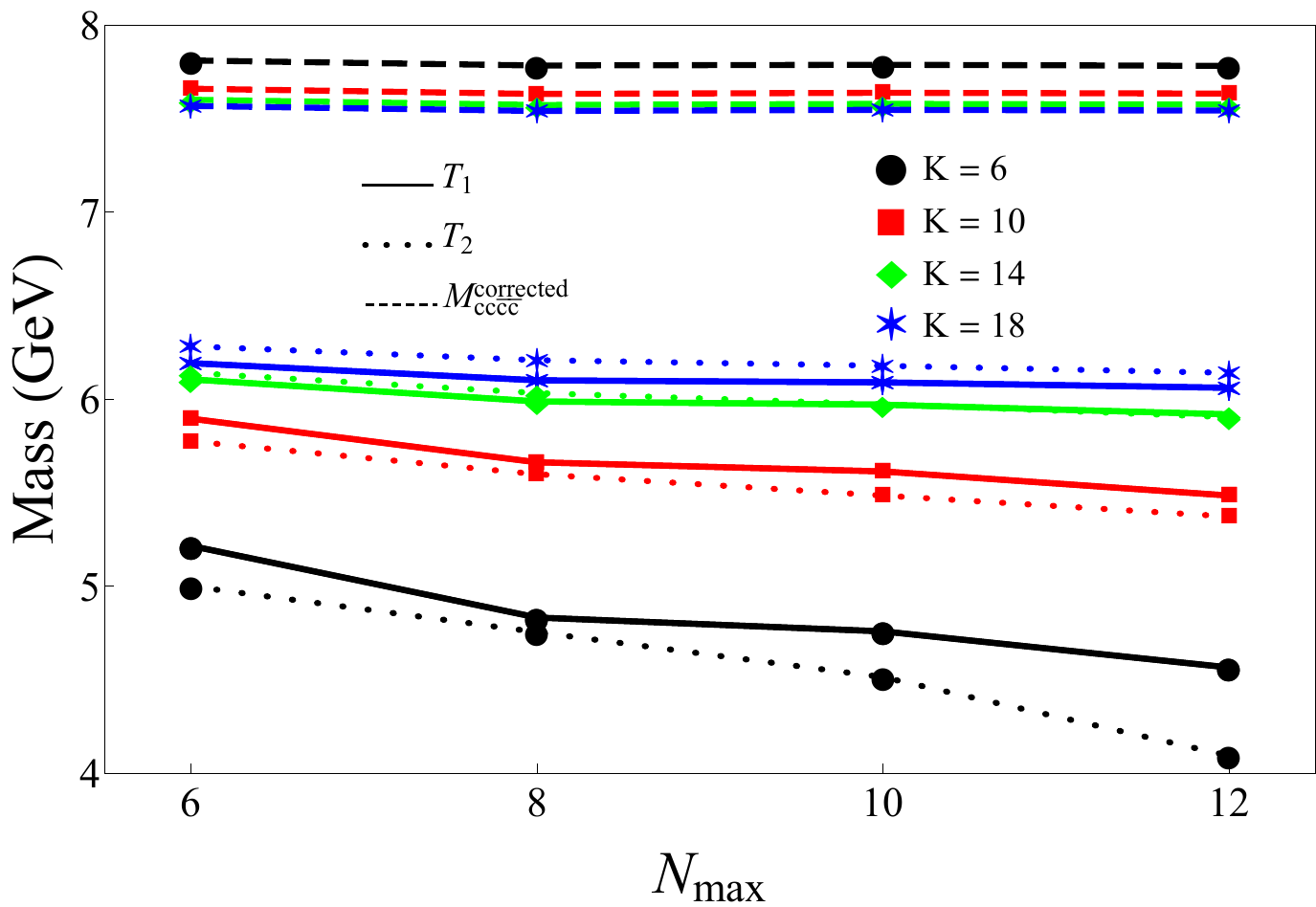}
 \caption{Threshold estimates $T_1$, $T_2$, and tetraquark
 masses $M^\mathrm{corrected}_{c c \bar c \bar c}$ depending
 on $N_\mathrm{max}$ and $K$. Dashed lines connect symbols
 representing $M^\mathrm{corrected}_{c c \bar c \bar c}$,
 solid lines connect symbols representing $T_1$, dotted
 lines connect symbols representing $T_2$. Different symbols
 represent different $K$. With increasing $K$ the threshold
 lines go up, while tetraquark lines go down.
 \label{fig:T1T2Mcorr}}
\end{figure}

By turning off interactions between particles
that do not belong to the same meson we can compute numerically
the invariant mass of two mesons occupying almost the
same finite basis -- we have to make identical particles
distinguishable because otherwise one would not be able
to consistently turn off, for example, an interaction
between $1$ and $4$ and at the same time keep interaction
between $1$ and $3$ turned on. Therefore, we define,
\beq
T_2
\es
\sqrt{
  M^2_{\text{two-meson}}(N_\mathrm{max},K)
- \Delta M^2(N_\mathrm{max},K)
}
\ ,
\label{eq:threshold2}
\eeq
where $M^2_\text{two-meson}$ is the ground state mass
in the aforementioned calculation of two-meson system
in a tetraquark calculation. Results for threshold
estimates and tetraquark masses are summarized in
Table~\ref{tab:results} and plotted in Fig.~\ref{fig:T1T2Mcorr}.
\begin{table*}[h]
	\begin{ruledtabular}
		\caption{Values (in MeV) of threshold estimates
		$T_1$, $T_2$, and corrected tetraquark masses
		$M^\mathrm{corrected}_{c c \bar c \bar c}$ for
		various $N_\mathrm{max}$ and $K$.
		\label{tab:results}}
		\begin{tabular}{ccccccccccccccccc}
		$K$
		& \multicolumn{4}{c}{6}
		& \multicolumn{4}{c}{10}
		& \multicolumn{4}{c}{14}
		& \multicolumn{4}{c}{18}
		\\
		$N_\mathrm{max}$
		& 6 & 8 & 10 & 12
		& 6 & 8 & 10 & 12
		& 6 & 8 & 10 & 12
		& 6 & 8 & 10 & 12
		\\
		\hline
		$T_{1}$
		& $5215$ & $4832$ & $4758$ & $4565$
		& $5895$ & $5662$ & $5613$ & $5484$
		& $6105$ & $5987$ & $5970$ & $5918$
		& $6192$ & $6100$ & $6089$ & $6060$
		\\
		$T_{2}$
		& $4999$ & $4754$ & $4513$ & $4093$
		& $5774$ & $5598$ & $5484$ & $5372$
		& $6140$ & $6032$ & $5972$ & $5903$
		& $6282$ & $6208$ & $6178$ & $6140$
		\\
		$M^\mathrm{corrected}_{c c \bar c \bar c}$
		& $7810$  & $7783$ & $7787$ & $7781$
		& $7659$  & $7631$ & $7637$ & $7633$
		& $7600$  & $7572$ & $7578$ & $7574$
		& $7567$  & $7540$ & $7546$ & $7542$
		\end{tabular}
	\end{ruledtabular}
\end{table*}

\begin{figure}
 \centering
 \includegraphics[width=\linewidth]{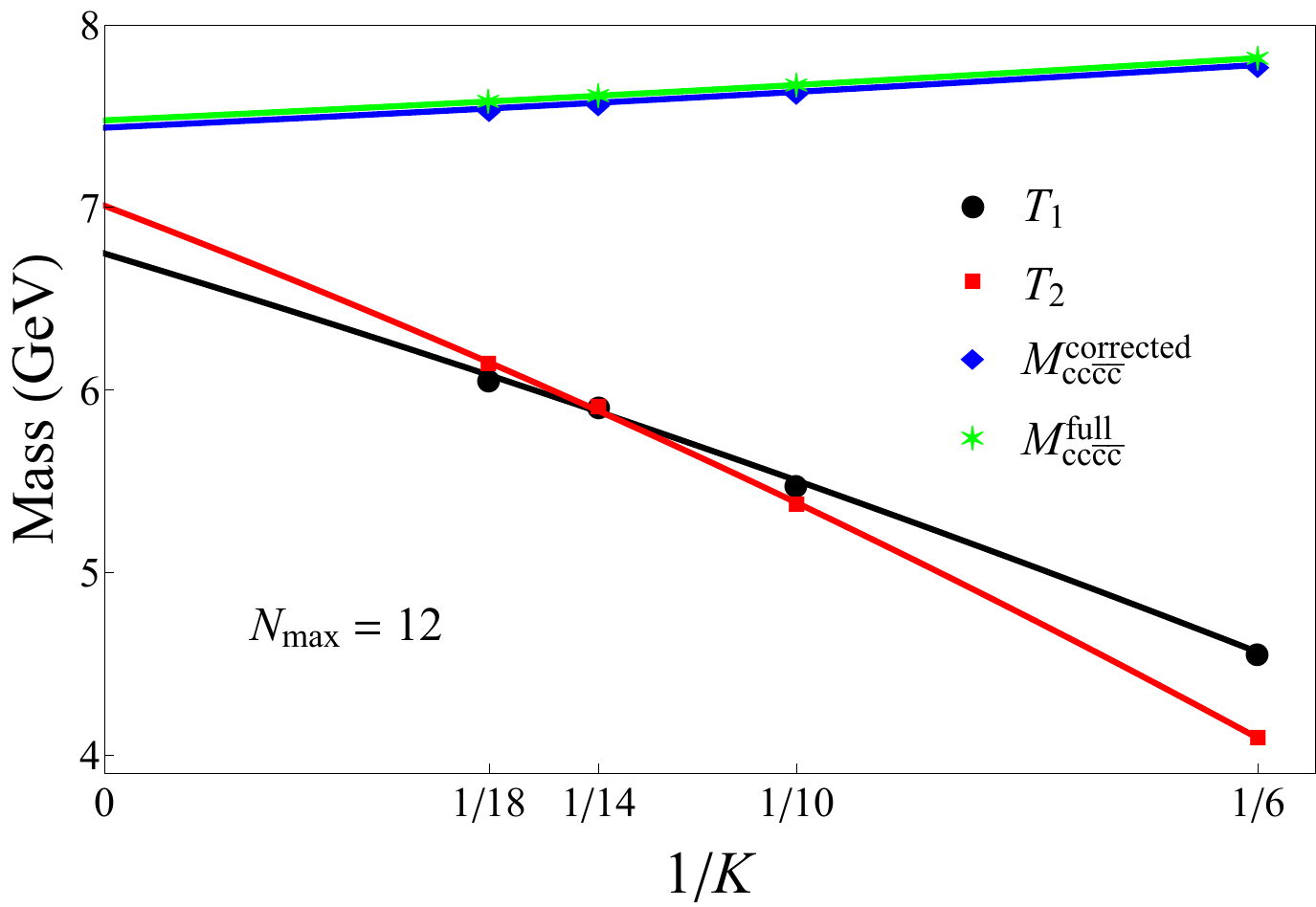}
 \caption{Threshold estimates and tetraquark masses as
 functions of $1/K$ for $N_\mathrm{max}$ = 12. The fitted
 solid lines (of the form $a + b/K + c/K^2$) are used to
 extrapolate the results to the point $1/K$ = 0.
 \label{fig:extrapolation}}
\end{figure}

Figure~\ref{fig:extrapolation} shows the result of
least squares fit of $a + b/K + c/K^2$ to threshold
estimates and tetraquark masses for $N_\mathrm{max}$ = 12.
The results for parameter $a$, \textit{i.e.}, extrapolations
of the fitted curves to the point $1/K$ = 0 are
$T_1$ = (6748~$\pm$~225)~MeV, $T_2$ = (7009~$\pm$~111)~MeV,
$M^\mathrm{full}_{c c \bar c \bar c}$ = (7477~$\pm$~2)~MeV,
$M^\mathrm{corrected}_{c c \bar c \bar c}$ = (7438~$\pm$~2)~MeV.
All those numbers are expected to go down in the limit
$N_\mathrm{max} \to \infty$ (provided we do not refit our
meson masses), but we expect that the shift should be much
smaller than the shift due to $K \to \infty$ extrapolation.
Our gluon mass introduces additional shift upwards on the
order of the value of $\mu$, \textit{i.e.}, 10~MeV. All
tetraquark masses lie substantialy above all threshold
estimates, including the extrapolations. These results
indicate that the lowest $c c \bar c \bar c$ eigenstate of
our model Hamiltonian is not bound with respect to breakup
into two separated mesons. It could be a resonant state.
However, such a conclusion would require additional
confirmation in the form of decay analysis.

\section{Conclusion}
\label{sec:conclusion}

We have done the first, to our knowledge, study of
all-heavy tetraquark states using a Hamiltonian in
the Front Form of dynamics, where all quarks are treated
individually, color degrees of freedom are unconstrained
(apart from the restriction to global color singlets)
and antisymmetrizations due to identical particles are
taken into account.

We note, however, that our confining potential breaks
the cluster decomposition principle, but the breaking
should be rather small for a nonrelativistic system
like all-charm tetraquark. Attempts to restore it
exactly lead to unphysical states with negative mass
squared. Therefore, our confining potential should be
regarded as an approximate effective potential with
a limited range of applicability.

Even without the negative $M^2$ problem,
confining long-range forces lead to problematic long-range
van der Waals forces~\cite{Greenberg:1981xn,Liu:1983rv}.
Such long-range forces are unlikely to be present in QCD.
A more probable picture would involve effective, massive
gluons to be the source of confining forces. They may or
may not form strings, but in any case a force mediated
by such gluons would be short-ranged.

All our estimates for the $c c \bar c \bar c$ ground state
mass turn out to be substantially higher than the estimates
we made for the lowest threshold for breakup into two 
$c \bar c$ mesons. Therefore, in our model, the ground state
tetraquark is unstable against dissociation into two charmonia.
There remain, however, open questions. For example, what would
happen if we used much larger basis spaces? Our estimates seem
to indicate a gap between the two-meson threshold and the lowest
tetraquark, but ultimately, close to the threshold, we should
see a lot of states filling a continuum spectrum. We might
also see some molecular states bound by the van der Waals
forces. The 15\% color-independent admixture might play
a role here, because it is, perhaps a bit counterintuitively,
likely to work against binding of tetraquark states.
This is because all pairs of quarks in a tetraquark
contribute an upward shift of mass due to zero-point energy,
while for the OGE-like case four of the potentials cancel
each other to a large extent in the color configuration
with two color singlets. At the same time, by design, meson
spectra and two-meson spectra are unaffected by the admixture.

Even if the ground state tetraquark is unstable, there
may still be stable tetraquarks higher in the mass, because
their thresholds can be higher. For example, authors of
Ref.~\cite{Barnea:2006sd}, using hyperspherical expansion
to solve Schr\"odinger equation, find exotic states $0^{+-}$
(6515~MeV) and $2^{+-}$ (6586~MeV) to be substantially
below their respective thresholds for dissociation.

It is also worth noting that the results for tetraquark
masses seem to be much more reliable than the threshold
estimates that we obtain, as can be seen in
Fig.~\ref{fig:extrapolation} and from our extrapolations.
Fit uncertainties are very small for tetraquarks, while
very large for threshold estimates. This is fortunate,
because meson calculations require far less amount of
computational
resources, hence, can be straightforwardly improved.
Therefore, we could fit parameters using the extrapolations
$K \to \infty$ and $N_\mathrm{max} \to \infty$ of mesons
masses (instead of at fixed $K$ and $N_\mathrm{max}$).
This would give us the threshold at physical values, while
extrapolations of tetraquark masses from, comparatively,
smaller $K$ and $N_\mathrm{max}$ would still give reliable
results.

\begin{acknowledgments}
K. S. is supported by the Chinese Academy of Sciences
President's International Fellowship Initiative (PIFI),
Grant No. 2021PM0066, the Chinese Ministry of Science
and Technology, Foreign Expert Project, Grant No. QN20200143003
and National Natural Science Foundation of China (NSFC)
under Grant No. 12047555. Z. K. and X. Z. are supported
by new faculty startup funding by the Institute of Modern
Physics, Chinese Academy of Sciences, by Key Research
Program of Frontier Sciences, Chinese Academy of Sciences,
Grant No. ZDB-SLY-7020, by the Natural Science Foundation
of Gansu Province, China, Grant No. 20JR10RA067, by the Foundation for Key Talents of Gansu Province, and by
the Strategic Priority Research Program of the Chinese
Academy of Sciences, Grant No. XDB34000000. 
This material is based upon work supported by the U.S. Department of Energy, Office of Science, under Award
Numbers DE-FG02-87ER40371 and DE-SC0018223 (SciDAC4/NUCLEI).
This research used computing resources of Gansu Computing
Center and Gansu Advanced Computing Center.
\end{acknowledgments}

\appendix*

\section{Antisymmetrization of the basis and color projection}

The color space for two quarks and two antiquarks is
$3^4 = 81$-dimensional. Assuming color confinement,
only color-singlet states can be physically realized.
The space of $Q Q \bar Q \bar Q$ color-singlet states
is only two-dimensional. It is, therefore, worth working
with color-singlet states only, because it means that
the matrices that need to be diagonalized numerically
have, roughly speaking, 40 times smaller dimensions.
We refer the reader to Fig.~12 in Ref.~\cite{Vary:2009gt}
for detailed examples of numbers of color singlets in
sectors with more particles. One has to, however, invest 
extra effort in the evaluation of the matrix
elements of the Hamiltonian.

The first step, which needs to be done in any case,
is to define a space of states with arbitrary color
that takes into account that some particles are identical.
One can use states
\beq
\ket{1234}
\es
B_1^\dag B_2^\dag D_3^\dag D_4^\dag \ket{0}
\ .
\eeq
Each particle is characterized by five quantum numbers:
$k_i$ -- longitudinal momentum number,
$n_i$ -- transverse-harmonic-oscillator radial number,
$m_i$ -- transverse-harmonic-oscillator angular number,
$\sigma_i$ -- light-front helicity, $c_i$ -- color.
$i = 1$, $2$, $3$, $4$. For each state $\ket{1234}$
there exist several other states that are linearly
dependent, \textit{e.g.}, $\ket{2134} = - \ket{1234}$
and $\ket{1243} = - \ket{1234}$. Moreover, some states
are identically zero, \textit{e.g.}, $\ket{1134} = 0$.
To define a proper orthonormal basis one has to
constrain possible quantum numbers of the particles.
Since quarks are fermions this can be done using
a relation of strict order. We say that $1 > 2$ if and
only if $k_1 > k_2$, or $k_1 = k_2$ and $n_1 > n_2$, or
$k_1 = k_2$ and $n_1 = n_2$ and $m_1 > m_2$, or
$k_1 = k_2$ and $n_1 = n_2$ and $m_1 = m_2$ and
$\sigma_1 > \sigma_2$, or
$k_1 = k_2$ and $n_1 = n_2$ and $m_1 = m_2$ and
$\sigma_1 = \sigma_2$ and $c_1 > c_2$. We define
our basis to contain only such states $\ket{1234}$
for which $1 > 2$ and $3 > 4$. Taking another such
state $\ket{1'2'3'4'}$ with $1' > 2'$ and $3' > 4'$
we have $\braket{1234}{1'2'3'4'} = \delta_{11'} \delta_{22'}
\delta_{33'} \delta_{44'}$, where $\delta_{ij} =
\delta_{k_i,k_j} \delta_{n_i,n_j} \delta_{m_i,m_j}
\delta_{\sigma_i,\sigma_j} \delta_{c_i,c_j}$.

In the second step we need to find color-singlet states,
which are defined as the kernel of the quadratic Casimir
operator $C_2 = \sum_{a=1}^8 \hat T^a \hat T^a$, where
\beq
\hat T^a
\es
\sum_{12}
\delta_{k_1, k_2} \delta_{n_1, n_2} \delta_{m_1, m_2}
\delta_{\sigma_1, \sigma_2}
\left(
  t^a_{c_1c_2} B_1^\dag B_2
- t^a_{c_2c_1} D_1^\dag D_2
\right)
\ ,
\eeq
where $\sum_{12}$ is the sum over all quantum numbers of
particles $1$ and $2$. We omit the part for gluons,
because we do not have gluons in our model. The color
operators $\hat T^a$ do not change any of the momentum
and spin quantum numbers, hence, $C_2$ is diagonal in
momentum and spin. Therefore, we can separately diagonalize
$C_2$ in subspaces of fixed momentum and spin quantum
numbers. Note that our relation $i > j$ compares colors
$c_i$ and $c_j$ in the very end, only if all other quantum
numbers turned out to be the same. 

There are four different
kinds of subspaces. To classify them it is convenient
to introduce another two relations. We say that
$i \approx j$ if all quantum numbers of $i$ and $j$
except color are the same (colors can be arbitrary).
We say that $i \gg j$ if $i > j$ and not $i \approx j$.
In other words, either $k_i > k_j$ or $k_i = k_j$ and
$n_i > n_j$ or $k_i = k_j$ and $n_i = n_j$ and $m_i > m_j$
or $k_i = k_j$ and $n_i = n_j$ and $m_i = m_j$ and
$\sigma_i > \sigma_j$. Hence, $\gg$ is just like $>$
except it does not take into account color. We can now
easily classify the four cases of color spaces.

Case 1: $1 \gg 2$ and $3 \gg 4$. In this case all 81 color
combinations are allowed. The color-singlet subspace is
two-dimensional and spanned by,
\beq
\ket{1234, S}
\es
\frac{1}{2\sqrt{6}}
\Big(
  2 \ket{r r \bar r \bar r}
+ 2 \ket{g g \bar g \bar g}
+ 2 \ket{b b \bar b \bar b}
\nn&&
+ \ket{r g \bar r \bar g}
+ \ket{g r \bar r \bar g}
+ \ket{g r \bar g \bar r}
+ \ket{r g \bar g \bar r}
\nn&&
+ \ket{g b \bar g \bar b}
+ \ket{b g \bar g \bar b}
+ \ket{b g \bar b \bar g}
+ \ket{g b \bar b \bar g}
\nn&&
+ \ket{b r \bar b \bar r}
+ \ket{r b \bar b \bar r}
+ \ket{r b \bar r \bar b}
+ \ket{b r \bar r \bar b}
\Big)
\ ,
\label{eq:colorSinglet1}
\\
\ket{1234, A_1}
\es
\frac{1}{\sqrt{12}}
\Big(
  \ket{r g \bar r \bar g}
- \ket{g r \bar r \bar g}
+ \ket{g r \bar g \bar r}
- \ket{r g \bar g \bar r}
\nn&&
+ \ket{g b \bar g \bar b}
- \ket{b g \bar g \bar b}
+ \ket{b g \bar b \bar g}
- \ket{g b \bar b \bar g}
\nn&&
+ \ket{b r \bar b \bar r}
- \ket{r b \bar b \bar r}
+ \ket{r b \bar r \bar b}
- \ket{b r \bar r \bar b}
\Big)
\ ,
\label{eq:colorSinglet2}
\eeq
where kets on the right hand sides are denoted by colors
$\ket{c_1 c_2 c_3 c_4}$ and we omit momentum and spin
quantum numbers, which are the same for each ket. Instead
of $1$, $2$, $3$, colors are called $r$, $g$, $b$,
respectively, for quarks and $\bar r$, $\bar g$, $\bar b$,
respectively, for antiquarks. For completeness, $r < g < b$
and $\bar r < \bar g < \bar b$. Note that $\ket{1234, S}$
is symmetric for either $1 \leftrightarrow 2$ or
$3 \leftrightarrow 4$, while $\ket{1234, A_1}$ is
antisymmetric for either $1 \leftrightarrow 2$ or
$3 \leftrightarrow 4$.

Case 2: $1 \approx 2$ and $3 \gg 4$. Now only three
quark-quark color combinations, $bg$, $br$ and $gr$,
are allowed (because $1 > 2$ still holds) while the
colors of antiquarks are unconstrained. Therefore,
this color space is 27-dimensional and there is only
one, antisymmetric, color-singlet combination,
\beq
\ket{1234, A_2}
\es
\frac{1}{\sqrt{6}}
\Big(
  \ket{g r \bar g \bar r}
- \ket{g r \bar r \bar g}
+ \ket{b r \bar b \bar r}
\nn &&
- \ket{b r \bar r \bar b}
+ \ket{b g \bar b \bar g}
- \ket{b g \bar g \bar b}
\Big)
\ .
\label{eq:colorSinglet3}
\eeq

Case 3: $1 \gg 2$ and $3 \approx 4$. Analogously
to case 2 only three antiquark-antiquark color
combinations, $\bar b \bar g$, $\bar b \bar r$
and $\bar g \bar r$, are allowed, because $3 > 4$,
while the colors of quarks are unconstrained.
The color space is again 27-dimensional and there
is only one, antisymmetric, color-singlet combination,
\beq
\ket{1234, A_3}
\es
\frac{1}{\sqrt{6}}
\Big(
  \ket{g r \bar g \bar r}
- \ket{r g \bar g \bar r}
+ \ket{b r \bar b \bar r}
\nn &&
- \ket{r b \bar b \bar r}
+ \ket{b g \bar b \bar g}
- \ket{g b \bar b \bar g}
\Big)
\ .
\label{eq:colorSinglet4}
\eeq

Case 4: $1 \approx 2$ and $3 \approx 4$. Both quarks
and antiquarks have constrained colors, because $1 > 2$
and $3 > 4$. The color space is 9-dimensional and there
is only one, antisymmetric, color-singlet combination,
\beq
\ket{1234, A_4}
\es
\frac{1}{\sqrt{3}}
\left(\phantom{\frac{}{}}
  \ket{g r \bar g \bar r}
+ \ket{b r \bar b \bar r}
+ \ket{b g \bar b \bar g}
\phantom{\frac{}{}}\right)
\ .
\label{eq:colorSinglet5}
\eeq

The third and last step is to calculate the common
factors in Hamiltonian matrix elements between states
with different color-singlet configurations that arise
due to color and antisymmetrization. We summarize
the results. In general we need to evaluate matrix
elements of the following types of operators,
\beq
\hat V_q
\es
\sum_{5', 5}
\delta_{c_{5'}, c_5}
\ V_{5'; 5}
B_{5'}^\dag B_5
\ ,
\\
\hat V_{\bar q}
\es
\sum_{5', 5}
\delta_{c_{5'}, c_5}
\ V_{5'; 5}
D_{5'}^\dag D_5
\ ,
\\
\hat V_{qq}
\es
\sum_{5', 6', 5, 6}
C^{qq}_{5', 6'; 5, 6}
\ V_{5', 6'; 5, 6}
\ \frac{1}{2} \, B_{5'}^\dag B_{6'}^\dag B_6 B_5
\ ,
\\
\hat V_{\bar q \bar q}
\es
\sum_{5', 6', 5, 6}
C^{\bar q \bar q}_{5', 6'; 5, 6}
\ V_{5', 6'; 5, 6}
\ \frac{1}{2} \, D_{5'}^\dag D_{6'}^\dag D_6 D_5
\ ,
\\
\hat V_{q \bar q}
\es
\sum_{5', 6', 5, 6}
C^{q \bar q}_{5', 6'; 5, 6}
\ V_{5', 6'; 5, 6}
\ B_{5'}^\dag D_{6'}^\dag D_6 B_5
\ ,
\eeq
where $V_{5'; 5}$ and $V_{5', 6'; 5, 6}$ depend on all
quantum numbers except color, while $C^{qq}_{5', 6'; 5, 6}$,
$C^{\bar q \bar q}_{5', 6'; 5, 6}$ and
$C^{q \bar q}_{5', 6'; 5, 6}$ depend only on color. There are
two types of interactions: color-independent ones, for which,
\beq
C^{q q}_{5', 6'; 5, 6}
\es
C^{\bar q \bar q}_{5', 6'; 5, 6}
\rs
C^{q \bar q}_{5', 6'; 5, 6}
\rs
\delta_{c_{5'}, c_5} \, \delta_{c_{6'}, c_6}
\ ,
\eeq
and OGE-like interactions, for which,
\beq
C^{q q}_{5', 6'; 5, 6}
\es
\sum_{a = 1}^8 t^a_{5'5} t^a_{6'6}
\ ,
\\
C^{\bar q \bar q}_{5', 6'; 5, 6}
\es
\sum_{a = 1}^8 t^a_{55'} t^a_{66'}
\ ,
\\
C^{q \bar q}_{5', 6'; 5, 6}
\es
- \sum_{a = 1}^8 t^a_{5'5} t^a_{66'}
\ .
\eeq
Note, that $C^{q q}_{5', 6'; 5, 6} =
C^{\bar q \bar q}_{5', 6'; 5, 6}$. We label color
singlets with capital letters $I$ and $J$ that can
equal to $S$, $A_1$, $A_2$, $A_3$ or $A_4$.
The general matrix elements are,
\beq
\bra{1' 2' 3' 4', I}
\, \hat V_q \,
\ket{1 2 3 4, J}
\es
C^q_{IJ} \, S^q_I \, S^q_J \, V^q_{IJ}
\, \delta_{e_{3'}, e_3} \, \delta_{e_{4'}, e_4}
\ ,
\\
\bra{1' 2' 3' 4', I}
\, \hat V_{\bar q} \,
\ket{1 2 3 4, J}
\es
C^{\bar q}_{IJ} \, S^{\bar q}_I
\, S^{\bar q}_J \, V^{\bar q}_{IJ}
\, \delta_{e_{1'}, e_1} \, \delta_{e_{2'}, e_2}
\ ,
\\
\bra{1' 2' 3' 4', I}
\, \hat V_{qq} \,
\ket{1 2 3 4, J}
\es
C^{qq}_{IJ} \, S^q_I \, S^q_J \, V^{qq}_{IJ}
\, \delta_{e_{3'}, e_3} \, \delta_{e_{4'}, e_4}
\ ,
\\
\bra{1' 2' 3' 4', I}
\, \hat V_{\bar q \bar q} \,
\ket{1 2 3 4, J}
\es
C^{\bar q \bar q}_{IJ} \, S^{\bar q}_I
\, S^{\bar q}_J \, V^{\bar q \bar q}_{IJ}
\, \delta_{e_{1'}, e_1} \, \delta_{e_{2'}, e_2}
\ ,
\\
\bra{1' 2' 3' 4', I}
\, \hat V_{q \bar q} \,
\ket{1 2 3 4, J}
\es
C^{q \bar q}_{IJ}
\, S^{q \bar q}_I \, S^{q \bar q}_J
\, V^{q \bar q}_{IJ}
\ ,
\eeq
where $e_i$ stands for all quantum numbers of particle $i$
except color, \textit{i.e.}, $n_i$, $m_i$, $k_i$ and
$\sigma_i$. Hence, $\delta_{e_i, e_j}$ is a product
of four Kronecker deltas, $\delta_{k_i, k_j} \delta_{n_i, n_j}
\delta_{m_i, m_j} \delta_{\sigma_i, \sigma_j}$.
Color factors $C_{IJ}$ are given in
Table~\ref{tab:colorFactors}. Symmetry factors $S_I$
are given in Table~\ref{tab:symmetryFactors}.
Finally, $V$ factors are,
\beq
V^q_{SS}
\es
  V_{1'; 1} \ \delta_{e_{2'}, e_2}
- V_{1'; 2} \ \delta_{e_{2'}, e_1}
- V_{2'; 1} \ \delta_{e_{1'}, e_2}
+ V_{2'; 2} \ \delta_{e_{1'}, e_1}
\ ,
\\
V^q_{A_i A_j}
\es
  V_{1'; 1} \ \delta_{e_{2'}, e_2}
+ V_{1'; 2} \ \delta_{e_{2'}, e_1}
+ V_{2'; 1} \ \delta_{e_{1'}, e_2}
+ V_{2'; 2} \ \delta_{e_{1'}, e_1}
\ ,
\\
V^{\bar q}_{SS}
\es
  V_{3'; 3} \ \delta_{e_{4'}, e_4}
- V_{3'; 4} \ \delta_{e_{4'}, e_3}
- V_{4'; 3} \ \delta_{e_{3'}, e_4}
+ V_{4'; 4} \ \delta_{e_{3'}, e_3}
\ ,
\\
V^{\bar q}_{A_i A_j}
\es
  V_{3'; 3} \ \delta_{e_{4'}, e_4}
+ V_{3'; 4} \ \delta_{e_{4'}, e_3}
+ V_{4'; 3} \ \delta_{e_{3'}, e_4}
+ V_{4'; 4} \ \delta_{e_{3'}, e_3}
\ .
\\
V^{qq}_{SS}
\es
  V_{1', 2'; 1, 2} - V_{2', 1'; 1, 2}
- V_{1', 2'; 2, 1} + V_{2', 1'; 2, 1}
\ ,
\\
V^{qq}_{A_i A_j}
\es
  V_{1', 2'; 1, 2} + V_{2', 1'; 1, 2}
+ V_{1', 2'; 2, 1} + V_{2', 1'; 2, 1}
\ ,
\\
V^{\bar q \bar q}_{SS}
\es
  V_{3', 4'; 3, 4} - V_{4', 3'; 3, 4}
- V_{3', 4'; 4, 3} + V_{4', 3'; 4, 3}
\ ,
\\
V^{\bar q \bar q}_{A_i A_j}
\es
  V_{3', 4'; 3, 4} + V_{4', 3'; 3, 4}
+ V_{3', 4'; 4, 3} + V_{4', 3'; 4, 3}
\ ,
\\
V^{q \bar q}_{IJ}
\es
\big[
         V_{1', 3'; 1, 3}
+ a   \, V_{1', 4'; 1, 3}
+ a   \, V_{2', 3'; 1, 3}
+        V_{2', 4'; 1, 3}
\np
    b \, V_{1', 3'; 1, 4}
+ a b \, V_{1', 4'; 1, 4}
+ a b \, V_{2', 3'; 1, 4}
+   b \, V_{2', 4'; 1, 4}
\np
    b \, V_{1', 3'; 2, 3}
+ a b \, V_{1', 4'; 2, 3}
+ a b \, V_{2', 3'; 2, 3}
+   b \, V_{2', 4'; 2, 3}
\np
         V_{1', 3'; 2, 4}
+ a   \, V_{1', 4'; 2, 4}
+ a   \, V_{2', 3'; 2, 4}
+        V_{2', 4'; 2, 4}
\big]
\, \widetilde{\delta\delta}
\ ,
\label{eq:VqqbarIJ}
\eeq
where $a = -1$ if $I = S$ and $a = 1$ if $I = A_i$,
while $b = -1$ if $J = S$ and $b = 1$ if $J = A_j$.
$\widetilde{\delta\delta}$ stands for matching
Kronecker deltas in quantum numbers of the spectators
of the interaction and it is different for each of
the sixteen terms in Eq.~(\ref{eq:VqqbarIJ}).
For example, $V_{2', 4'; 1, 3}$ describes an interaction
where the final interacting quark-antiquark pair is
$2'4'$, while initial interacting quark-antiquark pair
is $13$. Hence, the final spectator quark-antiquark pair
is $1'3'$ and the initial spectator quark-antiquark pair
is $24$. Therefore, $\widetilde{\delta\delta} =
\delta_{e_{1'},e_2} \delta_{e_{3'},e_4}$ in this case.
$V^q$, $V^{\bar q}$, $V^{q q}$ and $V^{\bar q \bar q}$
need not to be defined for $I J = S A_j$ or $A_i S$,
because color factors are always zero in those cases.
It is also worth noting that if $I \in \{ A_3, A_4 \}$
and $J \in \{ S, A_1, A_2 \}$ or if $J \in \{ A_3, A_4 \}$
and $I \in \{ S, A_1, A_2 \}$, then $\delta_{e_{3'}, e_3}
\, \delta_{e_{4'}, e_4}$ is always zero. Similarly,
if  $I \in \{ A_2, A_4 \}$ and $J \in \{ S, A_1, A_3 \}$
or if $J \in \{ A_2, A_4 \}$ and $I \in \{ S, A_1, A_3 \}$,
then $\delta_{e_{1'}, e_1} \, \delta_{e_{2'}, e_2}$ is
always zero.

\begin{table}[h]
\begin{ruledtabular}
\caption{\label{tab:colorFactors}
Interaction color factors between color-singlet states.}
\begin{tabular}{ccccccccc}
\multicolumn{3}{c}{} &
\multicolumn{3}{c}{color indepen.} & \multicolumn{3}{c}{OGE-like}
\\
$IJ$ &
$C^q_{IJ}$ & $C^{\bar q}_{IJ}$ &
$C^{qq}_{IJ}$ & $C^{\bar q \bar q}_{IJ}$ & $C^{q \bar q}_{IJ}$ &
$C^{qq}_{IJ}$ & $C^{\bar q \bar q}_{IJ}$ & $C^{q \bar q}_{IJ}$
\\
\hline
$S S$ &
$1$ & $1$ &
$\frac{1}{2}$ & $\frac{1}{2}$ & $1$ &
$\frac{1}{6}$ & $\frac{1}{6}$ & $-\frac{5}{6}$
\\
$S A_j$ or $A_i S$ &
$0$ & $0$ &
$0$ & $0$ & $0$ &
$0$ & $0$ & $-\frac{1}{\sqrt{2}}$
\\
$A_i A_j$ &
$1$ & $1$ &
$\frac{1}{2}$ & $\frac{1}{2}$ & $1$ &
$-\frac{1}{3}$ & $-\frac{1}{3}$ & $-\frac{1}{3}$
\end{tabular}
\end{ruledtabular}
\end{table}

\begin{table}[h]
\begin{ruledtabular}
\caption{\label{tab:symmetryFactors}
Symmetry factors $S_I$ as functions of $I$.
Note that $S^{q \bar q}_I = S^{q}_I S^{\bar q}_I$.}
\begin{tabular}{cccccc}
$I$ & $S^{q}_I$ & $S^{\bar q}_I$ & $S^{q \bar q}_I$
\\
\hline
$S$ & $1$ & $1$ & $1$
\\
$A_1$ & $1$ & $1$ & $1$
\\
$A_2$ & $\frac{1}{\sqrt{2}}$ & $1$ & $\frac{1}{\sqrt{2}}$
\\
$A_3$ & $1$ & $\frac{1}{\sqrt{2}}$ & $\frac{1}{\sqrt{2}}$
\\
$A_4$ & $\frac{1}{\sqrt{2}}$ & $\frac{1}{\sqrt{2}}$ & $\frac{1}{2}$
\end{tabular}
\end{ruledtabular}
\end{table}

\bibliography{References.bib}
\end{document}